\documentclass[aps,prl,twocolumn,superscriptaddress, showkeys]{revtex4-1}

\usepackage{graphicx}
\usepackage{siunitx}
\usepackage{empheq}
\usepackage{amsmath}
\usepackage{amssymb}
\usepackage{txfonts}
\usepackage[colorlinks=true, linkcolor=blue,citecolor=red]{hyperref}

\renewcommand{\Re}{\operatorname{Re}}
\renewcommand{\Im}{\operatorname{Im}}
\renewcommand{\figurename}{Figure}
\newcommand{\im}{\mathrm{i}}
\newcommand{\e}{\mathrm{e}}

\newcommand{\rp}[1]{(\ref{#1})}

\newcommand{\abs}[1]{\left|{#1}\right|}

\newcommand{\av}[1]{\left\langle #1 \right\rangle}

\newcommand{\al}[1]{^{(#1)}}
\newcommand{\da}{^\dagger}

\newcommand{\pt}[1]{\left( #1 \right)}
\newcommand{\pq}[1]{\left[ #1 \right]}
\newcommand{\pg}[1]{\left\{ #1 \right\}}

\newcommand{\lpq}[1]{\left[ #1 \right.}

\newcommand{\rpq}[1]{\left. #1 \right]}

\newcommand{\ee}{{\rm e}}
\newcommand{\ii}{{\rm i}}

\newcommand{\nn}{{\nonumber}}

\newcommand{\mm}[2]{ \pt{
                      \begin{array}{cc}
                       #1 \\
                       #2
                     \end{array}  }  }

\newcommand{\GG}{{\cal G}}

\newcommand{\TT}{{\cal T}}

\begin{document}

\title{Enhancing sideband cooling by feedback--controlled light}
\author{Massimiliano Rossi}
	\affiliation{School of Higher Studies ``C. Urbani", University of Camerino, 62032 Camerino (MC), Italy}
	\affiliation{School of Science and Technology, Physics Division, University of Camerino, 62032 Camerino (MC), Italy}
\author{Nenad Kralj}
	\affiliation{School of Science and Technology, Physics Division, University of Camerino, 62032 Camerino (MC), Italy}
\author{Stefano Zippilli}
\author{Riccardo Natali}
	\affiliation{School of Science and Technology, Physics Division, University of Camerino, 62032 Camerino (MC), Italy}
	\affiliation{INFN, Sezione di Perugia, 06123 Perugia (PG), Italy}
\author{Antonio Borrielli}
	\affiliation{Institute of Materials for Electronics and Magnetism, Nanoscience-Trento-FBK Division, 38123 Povo (TN), Italy}
\author{Gregory Pandraud}
	\affiliation{Delft University of Technology, Else Kooi Laboratory, 2628 Delft, The Netherlands}
\author{Enrico Serra}
	\affiliation{Delft University of Technology, Else Kooi Laboratory, 2628 Delft, The Netherlands}
	\affiliation{Istituto Nazionale di Fisica Nucleare, TIFPA, 38123 Povo (TN), Italy}
\author{Giovanni Di Giuseppe}
	\email{gianni.digiuseppe@unicam.it}
\author{David Vitali}
	\email{david.vitali@unicam.it}
	\affiliation{School of Science and Technology, Physics Division, University of Camerino, 62032 Camerino (MC), Italy}
	\affiliation{INFN, Sezione di Perugia, 06123 Perugia (PG), Italy}
	\affiliation{CNR-INO, L.go Enrico Fermi 6, I-50125 Firenze, Italy}
\date{\today}

\begin{abstract}
We realise a phase--sensitive closed--loop control scheme to engineer the fluctuations of the pump field which drives an optomechanical system, and show that the corresponding cooling dynamics can be significantly improved.
In particular, operating in the 
counter--intuitive
``anti--squashing" regime of positive feedback and increased field fluctuations, sideband cooling of a nanomechanical membrane within an optical cavity can be improved by 7.5~dB with respect to the case without feedback. Close to the quantum regime of reduced thermal noise, such feedback--controlled light would allow going well below the quantum backaction cooling limit.
\end{abstract}

\maketitle

\begin{figure*}[htt!]
\begin{center}
     {\includegraphics[width=1\textwidth]{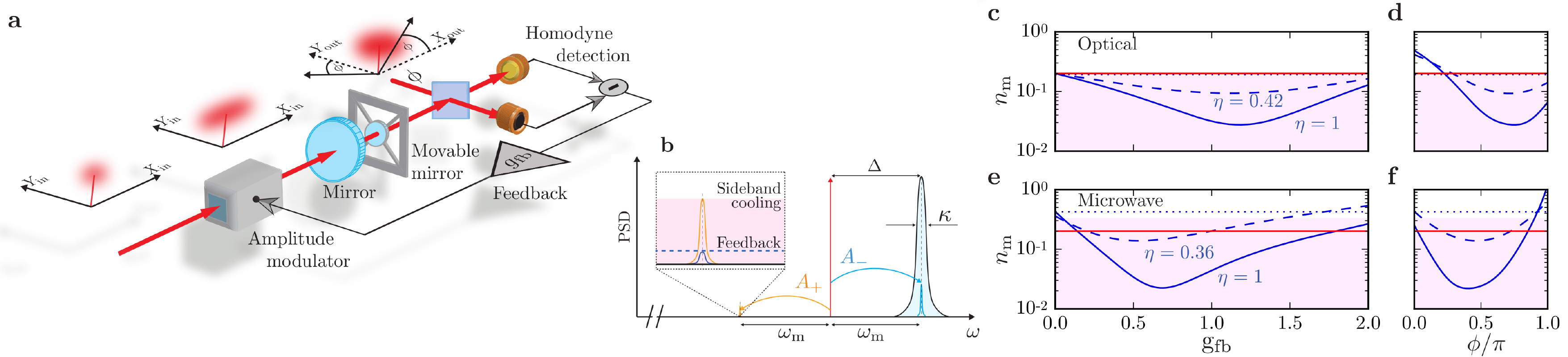}}
\caption{(Color online)
	{\bf a}, A cavity is driven by a coherent field with amplitude quadrature $\hat X_\mathrm{in}$ modified by an amplitude modulator fed with the output of the homodyne quadrature detection. 
	The output field quadratures, $\hat X_\mathrm{out}$ and $\hat Y_\mathrm{out}$, become correlated, depending on the homodyne phase $\phi$ and on the non--resonant cavity driving at detuning $\Delta$.
	{\bf b}, The radiation pressure interaction in an optomechanical cavity with decay rate $\kappa$ yields sidebands at mechanical frequencies $\pm\omega_\mathrm{m}$, that correspond to processes which enhance ($A_+$, yellow) and reduce ($A_-$, blue) mechanical energy.
	{\bf c}-{\bf f}, Theoretical results for the phonon number of the cooled resonator at low temperature (when standard sideband cooling is limited by backaction noise), as a function of the feedback gain amplitude $\mathrm{g}_\mathrm{fb}$ (see~\cite{SM}) (panels {\bf c} and {\bf e}) and the homodyne phase $\phi$ (panels {\bf d} and {\bf f}), for the parameters of the experiment of Ref.~\cite{Peterson2016Laser-Cooling-o} (panels {\bf c} and {\bf d}) and of Ref.~\cite{Clark} (panels {\bf e} and {\bf f}). 
	The light--pink areas indicate results beyond the backaction limit.  
	Solid--blue curves represent results for
	perfect detection efficiency $\eta = 1$, dashed--blue for $\eta = 0.42$~\cite{Purdy:2013ys} in panels {\bf c} and {\bf d},  and $\eta = 0.36$~\cite{Mallet:2011aa} in panels {\bf e} and {\bf f}, and finally dotted--blue curves for no feedback. Red lines are 	
	the best up--to--date results  
        obtained for systems operating at the quantum backaction limit, in the optical~\cite{Peterson2016Laser-Cooling-o} (with standard sideband cooling) and microwave~\cite{Clark} (where squeezing is employed to improve the system performance) regimes.
}
\label{fig:Figure_1}
\end{center}
\end{figure*}

Feedback loops based on real--time continuous measurements~\cite{WisemanBook} are commonly used for stabilisation purposes, and they have also been successfully applied to the stabilisation of quantum systems~\cite{Sayrin2011realtime, Riste2013, Wilson2015Measurement-bas}. Typically a system is continuously monitored 
and the acquired signal drives the actuator which in turn drives the system to the desired target.
Here we demonstrate a novel approach to closed--loop control in which 
the feedback acts on an additional control field which is used to drive the system of interest. In particular,
the actuator acts on the control field in order to engineer its phase and amplitude fluctuations. The resulting feedback--controlled in--loop field is then exploited to manipulate the system and improve its performance.
In--loop optical fields have been studied for decades both theoretically \cite{Shapiro1987Theory-of-light,Taubman1995Intensity-feedb,Wiseman1998In-Loop-Squeezi,Wiseman1999Squashed-states} and experimentally \cite{Buchler1999Suppression-of-,Sheard2005Experimental-de}. A lot of effort has been made to reduce (squash) the noise exhibited by the field fluctuations inside the loop. However, in--loop sub--shot--noise fluctuations cannot be recognised as squeezed below the vacuum noise level, for two different reasons: firstly, the free field commutation relations are no longer valid for time events separated by more than the loop delay--time, since in--loop fields are not free fields~\cite{Wiseman1999Squashed-states}; secondly, the corresponding out--of--loop fields exhibit super--shot--noise fluctuations~\cite{Shapiro1987Theory-of-light}. Nevertheless, useful applications of these fields have been proposed and realised, e.g. suppression of the radiation pressure noise \cite{Buchler1999Suppression-of-},  removal of classical intensity noise \cite{Sheard2005Experimental-de}, and atomic line narrowing \cite{Wiseman1998In-Loop-Squeezi}. The common basis of these works is the negative feedback regime.
Negative feedback has also been successfully employed in mechanical~\cite{Cohadon1999,Poggio2007Feedback-Coolin,Vinante2008Feedback-Coolin}, and cavity optomechanical systems~\cite{Wilson2015Measurement-bas}, where an electromagnetic field is used to probe a mechanical resonator, and in turn to control the feedback actuator, which acts directly on the mechanical oscillator.
Engineered light fluctuations in the form of squeezed light have also been used in optomechanical systems to improve
both the detection sensitivity~\cite{McKenzie2002Experimental-De,LIGO2013,Peano2015Intracavity-Squ,Clark2016Observation-of-} and the cooling efficiency~\cite{Schafermeier2016aa,Clark,Asjad2016Suppression-of-}.
In the present work we show that it is possible to manipulate, with a feedback system [see~\figurename~\ref{fig:Figure_1} (a)], the fluctuations of the laser field that drives an optomechanical system
to enhance optomechanical sideband cooling~\cite{AspelmeyerRMP,Teufel2011,Chan2011,Peterson2016Laser-Cooling-o}. Our analysis demonstrates the effectiveness of this approach in two very different parameter regimes, and shows that the light fluctuations can be properly adapted to reduce the effects of the dominant heating processes    
under very different physical situations.
At low temperature, when standard sideband cooling is limited by backaction noise, we show that the Stokes heating processes can be coherently suppressed by 
destructive interference so that the quantum backaction limit can be surpassed. At high temperature, when the performance of sideband cooling is restrained by thermal noise, the feedback can be operated close to instability in order to enhance inelastic light scattering processes and to improve the cooling rate.


A vibrational mode of a mechanical object coupled to a cavity field can be cooled by laser light when the cavity is resonant with anti--Stokes processes, whereby incident photons are scattered to higher frequencies, accompanied by a corresponding reduction in mechanical energy [see~\figurename~\ref{fig:Figure_1} (b)]. Residual Stokes processes, instead, heat the mechanical resonator.
The rates for Stokes ($A_+$) and anti--Stokes ($A_-$) processes determine the ultimate efficiency of the cooling process, such that in the absence of other sources of noise cooling is constrained by backaction noise, which sets the lower limit to mechanical excitations
$n_\mathrm{m}^0={A_+}/\pt{A_- - A_+}$.
The parameters $A_+$ and $A_-$ depend upon the fluctuations of the cavity light, which can be manipulated 
enclosing the cooling light in a feedback loop.
The feedback operates measuring a generic field quadrature
and using the detected signal to modulate the input amplitude quadrature $\hat X_{\rm in}$, while the conjugate phase quadrature $\hat Y_{\rm in}$ remains untouched. In particular, $\hat X_{\rm in}$ is modified according to the feedback relation
\begin{eqnarray}\label{Xin}
	\hat X_{\rm in}\to \frac{
	1}
					 {1- 2\,g_{\rm fb}(\omega)\,{\zeta}_{\rm out}\al{\phi}(\omega)}
			[
			\hat X_{\rm in}
			+2\,g_{\rm fb}(\omega)\,{\zeta}_{\rm out}\al{\phi+\pi/2}(\omega)\ \hat Y_{\rm in}
			]
\end{eqnarray}
where $g_{\rm fb}(\omega)$ is the electronic feedback transfer function and $\zeta_{\rm out}\al{\phi}(\omega)$ and $\zeta_{\rm out}\al{\phi+\pi/2}(\omega)$, defined in Supplemental Material~\cite{SM}, describe the response of the output field to the input amplitude and phase fluctuations respectively; i.e., in the absence of feedback the detected output field quadrature is $\hat X\al{\phi}_{\rm out}=\zeta_{\rm out}\al{\phi}(\omega)\,\, \hat X_{\rm in}+\zeta_{\rm out}\al{\phi+\pi/2}(\omega)\,\, \hat Y_{\rm in}$, and additional noise terms not relevant to the discussion are omitted both in Eq.~\eqref{Xin} and in the inline equation~\cite{SM}.

The corresponding mechanical scattering rates
\cite{SM},
\begin{align}\label{Afb}
	A_\pm &=
		\frac{G^2}{2\kappa}\abs{\chi_\mathrm{c}(\mp\omega_\mathrm{m})
		+\left[\frac{2\ g_{\rm fb}(\mp\omega_\mathrm{m})\ \zeta_{\rm c}^{(0)}(\mp\omega_\mathrm{m})}
			 {1-2 \,g_{\rm fb}(\mp\omega_\mathrm{m})\,{\zeta}_{\rm out}\al{\phi}(\mp\omega_\mathrm{m})}\ \ee^{\ii\,\phi}\right]^\ast}^2
\end{align}
are proportional to the square of the optomechanical coupling strength $G$, and
are given by the superposition of two contributions.
The first term is related to the standard sideband laser cooling, which can be expressed in terms of the susceptibility $\chi_\mathrm{c}(\omega)=2\kappa/\pq{\kappa+\ii(\Delta-\omega)}$ of a cavity with linewidth $\kappa$ and detuning $\Delta$.
The second term is instead determined by the feedback loop. The function $\zeta_\mathrm{c}^{(0)}(\omega)$ describes the response of the cavity field amplitude $\hat X$ to the input amplitude fluctuations, i.e. in the absence of feedback it can be expressed in terms of the input field as
$\sqrt{2\kappa}\,\hat X=\zeta_{\rm c}^{(0)}(\omega)\,\hat X_\mathrm{in}+\zeta_{\rm c}^{(\pi/2)}(\omega)\,\hat Y_\mathrm{in}$\,
(once more, additional noise terms are omitted).
It is 
important to note 
that the feedback term sums up coherently, and can be properly optimised to enhance the performance of sideband cooling. Specifically, 
Stokes processes can be  fully suppressed, $A_+=0$ (and therefore the backaction limit is surpassed), when the cavity and feedback contributions interfere destructively and cancel each other, which is achieved setting the feedback gain value to $2\,g_{\rm fb}(-\omega_{\rm m})=\chi_{\rm c}(-\omega_{\rm m})^*/{\pq{\zeta_{\rm out}\al{\phi}(-\omega_{\rm m})-\zeta_{\rm c}\al{0}(-\omega_{\rm m})\,\ee^{\ii\,\phi}  }}$.  
So far we have assumed perfect detection efficiency, meaning that all the light lost by the cavity is detected and employed in the loop. In practice, at finite detection efficiencies, Stokes processes cannot be fully suppressed.
Nevertheless, also in realistic cases 
a strong reduction of $A_+$ is observed, and 
this approach 
can
outperform the best up--to--date results  
obtained for systems operating at the quantum backaction limit, both in the optical~\cite{Peterson2016Laser-Cooling-o} and in the microwave~\cite{Clark} regime, as shown in \figurename~\ref{fig:Figure_1}(c)--(f).
%
These results correspond to situations in which thermal noise is so low that standard sideband cooling is essentially limited by backaction noise.
In general, thermal fluctuations, characterised by the number of thermal excitations $n^{\rm th}_\mathrm{m}$, compete with the effect of the cooling light to determine the stationary phonon occupancy $n_\mathrm{m} = (\gamma_\mathrm{m}\, n^{\rm th}_\mathrm{m} + \Gamma_\mathrm{opt}\, n_\mathrm{m}^{0})/(\gamma_\mathrm{m} +\Gamma_\mathrm{opt})$,
where $\gamma_\mathrm{m}$ and $\Gamma_\mathrm{opt} = (A_- - A_+)$ are the mechanical and optical damping rates, respectively. Hence, at high temperature
aiming at barely suppressing Stokes processes becomes ineffective. However, 
in this regime, the
effects of thermal noise can be strongly reduced by operating the feedback close to instability such that $\Gamma_{\rm opt}$ is increased to large values, at the expense of increasing the backaction limit $n_\mathrm{m}^0$.


\begin{figure*}[ht!]
\begin{center}
    {\includegraphics[width=1\textwidth]{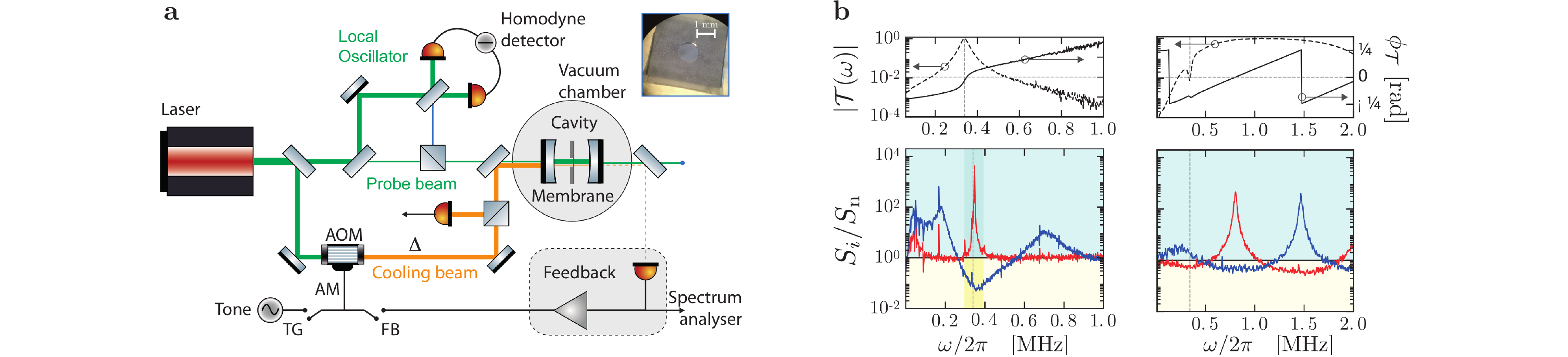}}
 \caption{(Color online)
	{\bf a}, Optomechanical cavity driven by the in--loop cavity mode (cooling beam).
	Dynamical backaction of the fundamental mechanical mode is provided by detuning the cooling beam by means of an acousto--optic modulator (AOM). Feedback is applied by amplitude modulating (AM) the cooling laser with an electronically processed copy of the transmitted photocurrent.
	Inset: image of the circular SiN membrane used, radius  $\SI{0.615}{\mm}$ and thickness $\SI{97}{\nm}$.
	The switches TG and FB allow the open-- and closed--loop transfer functions to be measured~\cite{SM}.
	A probe beam is used to monitor the cavity frequency fluctuations via balanced homodyne detection.
	{\bf b}, Current noise spectra $S_i$, normalised to the detection noise $S_\mathrm{n}$, of the transmitted (bottom--left) and reflected (bottom--right) photocurrent
	measured placing the membrane in a position of zero optomechanical interaction $g_0=0$. 
	Red and blue traces correspond to different signs of the amplifier output in the electronic filter.
	The vertical dashed grey line indicates the detuning $\Delta=2\pi\times\SI{330}{\kilo\hertz}$. The light--blue (light--yellow) area represents the anti--squashing (squashing) regime, where noise is amplified (reduced) below the detection noise. We refer to the gain for which the feedback based on transmission provides anti--squahing around the cavity detuning (shaded area) as positive.
	The top part shows the corresponding measured amplitude
	(dashed lines) and phase
	(solid lines) of the complex open--loop feedback response function $\mathcal{T}(\omega)$
	determined from the transmitted light detected with the switch FB open~\cite{SM}.
	}	
\label{fig:Figure_2}
\end{center}
\end{figure*}

We have tested this high temperature regime with a double--sided, \SI{90}{\milli\meter}--long symmetric cavity~\cite{Karuza:2012fk,Karuza2013aa}, with a decay rate $\kappa=2\pi\times\SI{20.15}{\kilo\hertz}$.
A $\mathrm{SiN}$ membrane is placed in the middle of the optical cavity~\cite{Thompson:2008uq,Karuza:2012fk}. It is a highly stressed circular membrane, with a diameter of \SI{1.2}{\milli\meter}, a thickness of \SI{97}{\nano\meter}, and 
negligible
optical absorption \cite{Serra2016Microfabricatio}. We focus on the fundamental mechanical mode, characterised by a resonance frequency  $\omega_\mathrm{m}=2\pi\times\SI{343.13}{\kilo\hertz}$ and a decay rate $\gamma_\mathrm{m}=2\pi\times\SI{1.18}{\hertz}$.
{The optomechanical coupling is}
$G=g_0 \sqrt{2\,n_c}$, with $n_c$ the number of cavity photons
and 
{$g_0$, the single--photon coupling,}
tunable 
by translating the membrane within the cavity standing wave~\cite{SM}.
Two beams, the probe (green lines) and the cooling beam (orange lines), are derived from a \SI{1064}{\nano\meter} master laser (see \figurename~\ref{fig:Figure_2}). The former, which is not part of the feedback loop, is used to lock the laser frequency to the cavity resonance, and to reveal the mechanical displacement by homodyne detection.
The cooling beam, detuned from the relevant cavity resonance by $\Delta=2\pi\times\SI{330}{\kilo\hertz}$
is, instead, enclosed in the feedback loop.  The amplitude quadrature
(corresponding to $\phi=0$)
of the transmitted (reflected) field is directly detected with a single photodiode and the resulting photocurrent, eventually filtered and amplified [in \figurename~\ref{fig:Figure_2}(a) the filter is applied to the transmitted light], is fed back to the input field by amplitude modulating (AM) the acousto--optic modulator (AOM)~\cite{SM}.

\begin{figure}[ht!]
\begin{center}
  {\includegraphics[width=.45\textwidth]{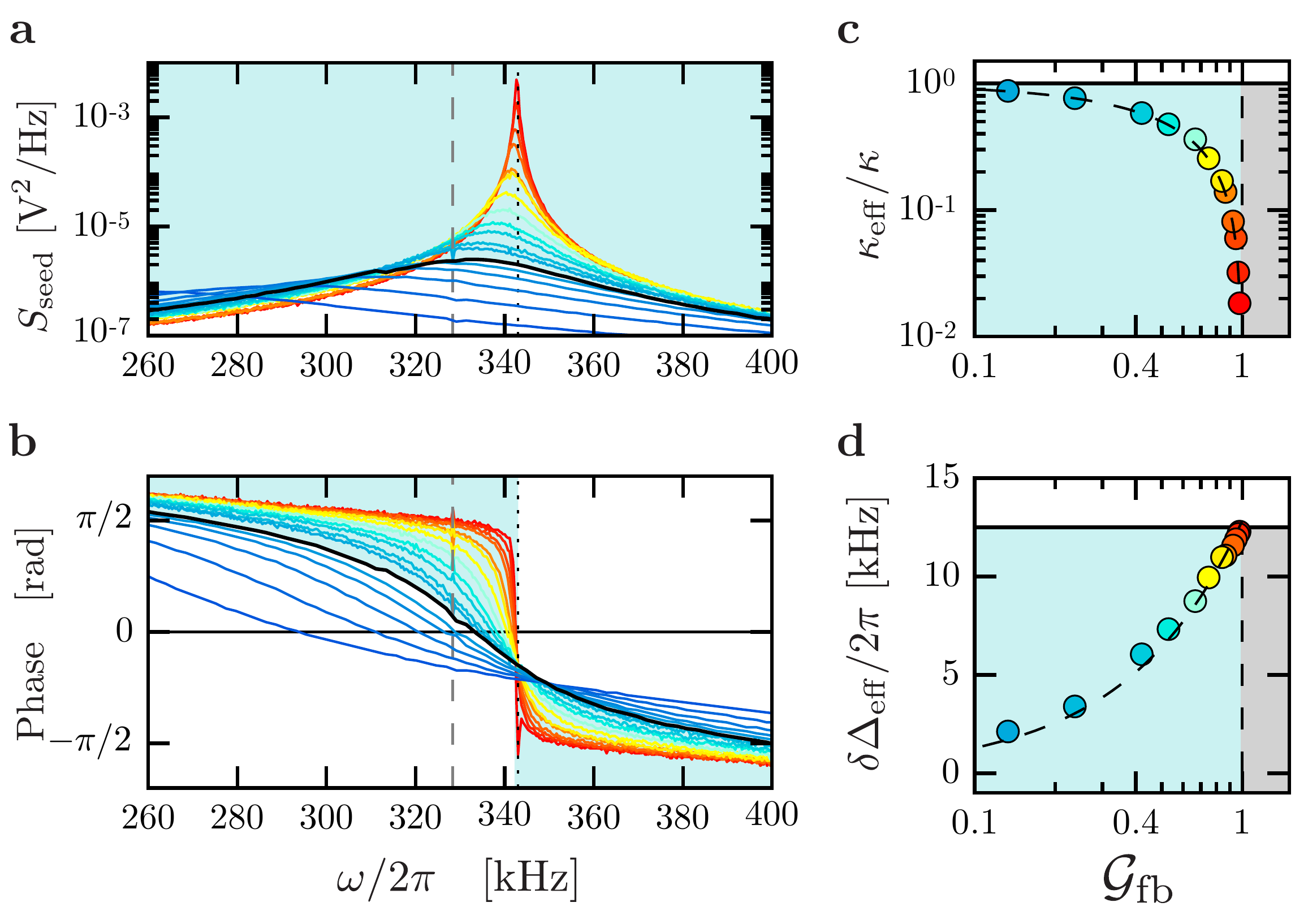}}
 \caption{(Color online)
 	{\bf a}, Amplitude and {\bf b}, phase of the measured closed--loop transfer function for the transmitted cooling field detuned by $\Delta = 2\pi\times\SI{330}{\kilo\hertz}$ (dashed grey line)~\cite{SM}. Black traces are acquired without the feedback loop. The dotted black line indicates the effective cavity frequency at instability.
	{\bf c}, Effective cavity decay rate $\kappa_\mathrm{eff}$, normalized to the out--of--loop decay rate $\kappa$, and {\bf d}, effective detuning shift $\delta\Delta_\mathrm{eff} \equiv \Delta_\mathrm{eff}-\Delta$ as a function of the positive feedback gain.
	For positive feedback (from light blue to red) the effective cavity decay rate $\kappa_{\mathrm{eff}}$ decreases and the effective detuning $\Delta_\mathrm{eff}$ increases, while for negative feedback (from light to dark blue) the situation is reversed. The grey area represents the instability region.
}
\label{fig:Figure_3}
\end{center}
\end{figure}

We first measure the in--loop light properties by placing the membrane at a node of the cavity field to rule out the optomechanical interaction. The feedback loop is fully characterised by measuring the open--loop transfer function $\mathcal{T}(\omega)$, which includes both the electronic and the optical response of the system [see \figurename~\ref{fig:Figure_2}(b), top, and~\cite{SM}].
The electronic part, ${g}_{\rm fb}(\omega)$, is generally complex due to the feedback delay--time $\tau_{\rm fb}$, which, in our case, is \SI{750}{\nano\second}. When the feedback loop is closed, the amplitude noise fluctuations are modified, as shown in \figurename~\ref{fig:Figure_2}(b), bottom;
the noise becomes frequency--dependent, with regions below (noise squashing) and above (noise anti--squashing) the noise level with no feedback~\cite{Shapiro1987Theory-of-light,Wiseman1999Squashed-states}.
The feedback--controlled cavity also shows a modified susceptibility in the anti--squashing regime, as can be verified by sending a weak classical seed field, larger than all noises, but too small to affect the mean cavity amplitude.
In the regime of our system, $\Delta\gg\kappa$ and small delay time $1/\tau_{\rm fb}\gg\kappa$
(so that a single anti-squashing resonance [see Fig.~\ref{fig:Figure_2} (b)] contributes to the dynamics), and for frequencies close to the cavity resonance, the seed experiences an effective cavity susceptibility
$\chi_\mathrm{c}^\mathrm{eff}(\omega)=2\kappa\,[\kappa_\mathrm{eff}+\im(\Delta_\mathrm{eff}-\omega)]^{-1}$,
with
$\kappa_{\mathrm{eff}}=\kappa\pt{1-\GG_{\rm fb}}$ and
$\Delta_{\mathrm{eff}}=\Delta-\kappa\,\GG_{\rm fb}\,\tan[\phi_\mathcal{T}(\Delta)]$,
where $\GG_{\rm fb}$ is  the normalised feedback gain, which is ${\GG_\mathrm{fb}}=1$ at the feedback stability threshold defined by $\kappa_\mathrm{eff}=0$, and $\phi_\mathcal{T}(\Delta)$ the phase of the feedback response function $\mathcal{T}(\omega)$ at the detuning $\Delta$~\cite{SM}.
%
%
%
%
%
%
Experimentally we determine
$\chi_\mathrm{c}^\mathrm{eff}(\omega)$ by measuring the closed--loop transfer function for different feedback gains (see \figurename~\ref{fig:Figure_3}). By increasing the gain the system approaches the feedback stability threshold, i.e. $\kappa_\mathrm{eff}$ tends to 0, as shown in \figurename~\ref{fig:Figure_3}(c)--(d).
We were able to reach a minimum effective cavity linewidth $\kappa_{\rm eff}\approx2\pi\times\SI{250}{\hertz}$ and a detuning $\Delta_\mathrm{eff}\approx2\pi\times\SI{342.5}{\kilo\hertz}$.

\begin{figure}[ht!]
\begin{center}
 \includegraphics[width=.45\textwidth]{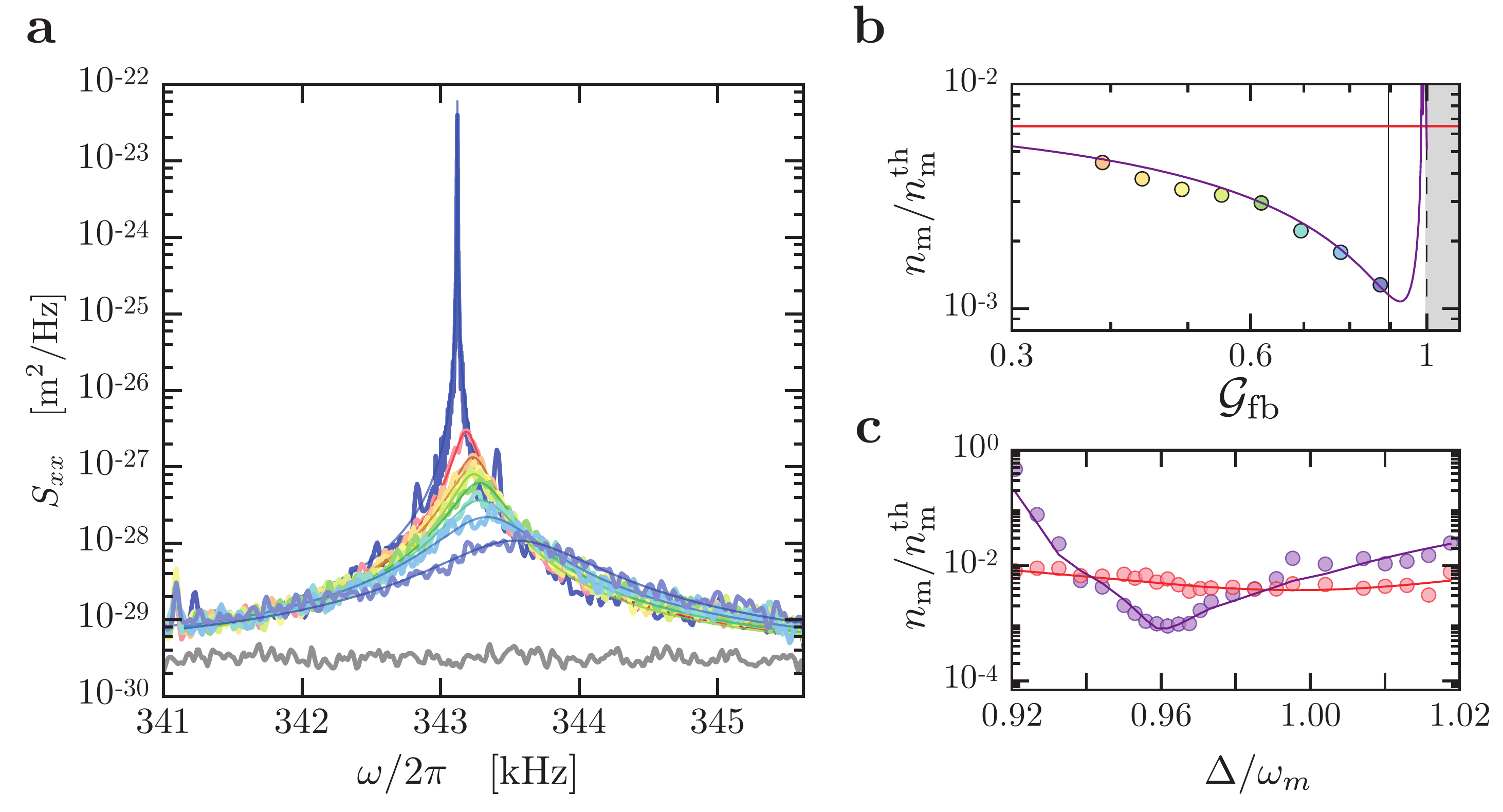}
 \caption{(Color online)
  	{\bf a}, Homodyne spectra of mechanical displacement noise $S_{xx}$.
	The blue trace represents the thermal fluctuations of the fundamental mechanical mode at \SI{300}{\kelvin}; the grey trace is the detection noise.
	 Dynamical backaction cools the mechanical motion down to \SI{2}{\kelvin}, as shown by the red trace (cooling beam on at \SI{33}{\micro\watt} and feedback off). From orange to light purple, the feedback is turned on and the gain increased.
	{\bf b} and {\bf c}, Effective energy reduction as a function of the gain 
	($\GG_{\rm fb}$) and normalised detuning 
	($\Delta/\omega_{\rm m}$)~\cite{SM}.
	Dots are experimental data. 
	Each dot in {\bf b} corresponds to the spectrum of the same colour in {\bf a}, the grey area represents the instability region, and the vertical grey line indicates the optimal gain value for cooling, used in plot {\bf c}.
	In {\bf c}, purple and red dots are results with and without feedback, respectively.
	In both panels purple and red lines are theoretical 
	results, computed using the measured parameters,
	with and without feedback, respectively.
	}
\label{fig:Figure_4}
\end{center}
\end{figure}

So far we have characterised the feedback system with $g_0=0$
and we have determined the properties of the cavity experienced by the resonator
which do not depend on the optomechanical interaction
(i.e. the values of $\kappa_\mathrm{eff}$ and $\Delta_\mathrm{eff}$).
These measurement are sufficient to perfectly reproduce the exprimental cooling results reported below.
The positive feedback regime (anti--squashing), which we focus on, enhances light amplitude noise. We show that driving an optomechanical cavity with the resulting in--loop field improves cooling by increasing the optical cooling rate.
The membrane is now placed in a position in which $g_0=2\pi\times\SI{0.84}{Hz}$~\cite{SM}.
Without feedback, a beam of \SI{33}{\micro\watt} red detuned by $\Delta=2\pi\times\SI{330}{\kilo\hertz}$ [see \figurename~\ref{fig:Figure_4}(a)] cools the membrane by
dynamical backaction~\cite{AspelmeyerRMP} from room temperature to an effective temperature of  \SI{2}{\kelvin}.
As the transmission feedback loop is closed and the gain is varied, the mechanical susceptibility, determined with the out-of-loop probe field, is shifted and broadened, as seen from~\figurename~\ref{fig:Figure_4}(a) and \cite{SM}.
The corresponding reduction of the effective mechanical energy reported in~\figurename~\ref{fig:Figure_4}(b)
in terms of the number of mechanical excitations $n_{\rm m}$,
and computed by numerical integration of the spectra~\cite{SM}, demonstrates an enhancement of the cooling rate, which settles the minimum effective temperature to 
$\hbar\omega_{\rm m}n_{\rm m}/k_B=\SI{350}{\milli\kelvin}$ for an optimal gain of $\mathcal{G}_\mathrm{fb}\sim 0.9$. Having fixed the optimal gain, we measured the effective mechanical energy as a function of the detuning [\figurename~\ref{fig:Figure_4}(c)], reaching the minimum phonon number for the optimal bare detuning $\Delta^\mathrm{opt}=2\pi\times\SI{329.4}{\kilo\hertz}$.
This value is consistent with the one estimated by using the measured feedback phase margin at the detuning frequency, $\phi_\mathcal{T}(\Delta^\mathrm{opt}) \sim \SI{-0.59}{\radian}$, and by setting
the feedback gain at instability, $\mathcal{G}_\mathrm{fb} = 1$, and the effective detuning at the optimal resolved sideband cooling condition $\Delta_\mathrm{eff} \sim \omega_\mathrm{m}$, that is $\Delta^\mathrm{opt} \sim \omega_\mathrm{m} + \kappa\,\tan\left[\phi_\mathcal{T}(\Delta^\mathrm{opt})\right]$.
%


Our results demonstrate that the in--loop field fluctuations obtained in the counter--intuitive
regime of positive feedback can be exploited for enhancing the cooling efficiency in an optomechanical system.
In the regime of low thermal noise we find theoretically that our approach allows the backaction limit to be beaten by almost an order of magnitude, both in the optical and the microwave regime [see \figurename~\ref{fig:Figure_1}(c)--(f)], as a result of the engineered intracavity field fluctuations, which contribute to the coherent cancellation of Stokes processes, and hence to the reduction of the backaction limit.
This result is analogous to that discussed in~\cite{Clark,Asjad2016Suppression-of-}, which makes use of squeezed light, but is achieved with a significantly simpler setup which does not require quantum nonlinearities.
The foreseen dynamics should be observable 
{including the feedback system in,} 
for example, the experimental setup of Refs.~\cite{Clark,Peterson2016Laser-Cooling-o}.
This would require using homodyne detection, with a properly optimised phase of the detected quadrature, and sufficiently large detection efficiency.
With our setup, which instead does not work at cryogenic temperature, we 
operate the feedback close to the instability and achieve
an enhancement of the cooling rate of 
\SI{10}{\decibel}, with a corresponding reduction of the phonon number of 
\SI{7.5}{\decibel} 
(see \figurename~\ref{fig:Figure_4}). 
In this limit the effectiveness of the feedback is associated with a reduction of the cavity linewidth, which increases the optomechanical cooperativity. The fact that the enhancement of the cooling rate is not reflected in an equal reduction of the number of mechanical excitations is due to the concomitant increase
of the backaction limit in this high temperature regime, which prevents further cooling of the resonator~\cite{SM}.
The generic technique that we have demonstrated can be adopted in a broad range of applications, whenever a system of interest is controlled with
an electromagnetic field subject to a phase--sensitive measurement, in the classical as well as the quantum regime.


\acknowledgments{
We acknowledge the support of the European Union's Horizon 2020 research and innovation program under grant agreement No 732894 (FET Proactive HOT).}


\onecolumngrid

\newpage

\setcounter{secnumdepth}{3}

\setcounter{equation}{0}
\setcounter{figure}{0}
\setcounter{section}{0}


\hspace{-.5cm}
\begin{minipage}{18.2cm}

\begin{center}
{\bf\large  Supplemental material for 
``Enhancing sideband cooling with  feedback--controlled light''
}

\bigskip 
{
Massimiliano Rossi,$^{1,2}$ Nenad Kralj,$^2$ Stefano Zippilli,$^{2,3}$ Riccardo Natali,$^{2,3}$ Antonio Borrielli,$^4$\\
Gregory Pandraud,$^5$ Enrico Serra,$^{5,6}$ Giovanni Di Giuseppe,$^{2,3}$ David Vitali,$^{2,3}$}

{\it \small
$^1$School of Higher Studies ``C. Urbani", University of Camerino, 62032 Camerino (MC), Italy\\
$^2$School of Science and Technology, Physics Division, University of Camerino, 62032 Camerino (MC), Italy\\
$^3$INFN, Sezione di Perugia, 06123 Perugia (PG), Italy\\
$^4$Institute of Materials for Electronics and Magnetism, Nanoscience-Trento-FBK Division, 38123 Povo (TN), Italy\\
$^5$Delft University of Technology, Else Kooi Laboratory, 2628 Delft, The Netherlands\\
$^6$Istituto Nazionale di Fisica Nucleare, TIFPA, 38123 Povo (TN), Italy
}

\smallskip
(Dated: \today)
\end{center}

\end{minipage}

\vspace{1cm}
\twocolumngrid

\section{Theory}
\label{sec:theor}

Here we introduce the general model at the basis of the findings reported in the Letter, and discuss
additional complementary theoretical results valid in the low noise limit.

The system dynamics can be analysed in terms of quantum Langevin equations for the evolution of annihilation and creation operators of cavity photons $\hat a$ and $\hat a\da$, and phononic excitations of the mechanical mode $\hat b$ and $\hat b\da$.
We assume the usual linearised regime of optomechanics,
where the operators describe fluctuations around steady state average values~\cite{Genes2008Ground-state-co_SM}.
It is therefore convenient to analyse it in Fourier space.
The corresponding equations are
\begin{eqnarray}\label{QLE}
-\ii\,\omega\, \hat b(\omega) &=&-\pt{\frac{\gamma_\mathrm{m}}{2}+\ii\,\omega_\mathrm{m}}\,\hat b(\omega)+\ii\,G\pq{\hat a(\omega)+\hat a\da(\omega)}
\\&&+\sqrt{\gamma_\mathrm{m}}\, \hat b_{\rm in}(\omega)
\nn\\
-\ii\,\omega\, \hat a(\omega)&=&-\pt{\kappa+\ii\,\Delta}\,\hat a(\omega)+\ii\,G\pq{\hat b(\omega)+\hat b\da(\omega)}
\nn\\&&\hspace{-1cm}
+\sqrt{2\kappa_0}\,\hat a_{\rm in,0}(\omega)\, \ee^{-\ii\theta_\Delta}+\sqrt{2\kappa_1}\,\hat a_{\rm in,1}(\omega)+\sqrt{2\kappa'}\,\hat a_{\rm in}'(\omega)\ ,\nn
\end{eqnarray}
where $\Delta$ is the detuning between cavity resonance and the laser frequency, $\pq{\hat a(\omega)}\da=\hat a\da(-\omega)$ (and analogously for other creation and annihilation operators), $\theta_\Delta=\arctan\pt{-\Delta/\kappa}$ is the phase shift of the cavity field with respect to the input due to the non-resonant driving, and $b_{\rm in}(\omega)$ is the environmental noise operator directly coupled to the mechanical resonator with rate $\gamma_\mathrm{m}$, characterised by an average number of thermal excitations $n_{\rm m}^{\rm th}$, such that $\av{\hat b_{\rm in}\da(\omega)\ \hat b_{\rm in}(\omega')}=n_{\rm m}^{\rm th}\ \delta(\omega+\omega')$. We have included several dissipation channels for the cavity, with corresponding vacuum noise operators $\hat a_{\rm in,0}(\omega)$, $\hat a_{\rm in,1}(\omega)$ and $\hat a_{\rm in}'(\omega)$, which are coupled to the system with rates $\kappa_0$, $\kappa_1$ and $\kappa'$ respectively, and such that the total cavity decay rate is $\kappa=\kappa_0+\kappa_1+\kappa'$.
This allows one to analyse different scenarios in which the feedback works by measuring either the reflected field (the feedback acts on the same output channel which is measured) or the transmitted one (detection and actuation are applied on different channels). This model describes, for example, a Fabry-P\'{e}rot cavity made of two mirrors with decay rates $\kappa_0$ and $\kappa_1$ respectively. Instead $\kappa'$ stands for additional internal losses.

The cavity output, either from the first or the second mirror, is detected and used to modulate the amplitude of the input through the first mirror, i.e. given the detected photocurrent $\hat i_{\rm fb}(\omega)$ (the specific form of which is given below), the input
operator is
\begin{eqnarray}
\hat a_{\rm in,0}(\omega)=\hat a_{\rm in,0}^\circ(\omega)+g_{\rm fb}(\omega)\ \hat i_{\rm fb}(\omega)\ ,
\end{eqnarray}
where $\hat a_{\rm in,0}^\circ$ is the input with no feedback and $g_{\rm fb}(\omega)$ is the feedback transfer function for which $g_{\rm fb}(\omega)^*=g_{\rm fb}(-\omega)$. This corresponds to a modulation of the amplitude quadrature $\hat X_{\rm in,0}(\omega)=\hat a_{\rm in,0}(\omega)+\hat a_{\rm in,0}\da(\omega)=\hat X_{\rm in,0}^\circ(\omega)+2\,g_{\rm fb}(\omega)\, \hat i_{\rm fb}(\omega)$ (where $\hat X_{\rm in,0}^\circ(\omega)$ is the input field amplitude with no feedback), while the input phase quadrature $\hat Y_{\rm in,0}(\omega)=-\ii\,\hat a_{\rm in,0}(\omega)+\ii\,\hat a_{\rm in,0}\da(\omega)$ remains unaffected by the feedback.
The output fields are given by the standard input-output relations
\begin{eqnarray}
\hat a_{\rm out,0}(\omega)&=&\sqrt{2\kappa_0}\ \hat a(\omega)\ \ee^{\ii\pt{\theta_\Delta-\bar\theta_\Delta}}-\hat a_{\rm in,0}(\omega)\ \ee^{-\ii\bar\theta_\Delta}
\nn\\
\hat a_{\rm out,1}(\omega)&=&\sqrt{2\kappa_1}\,\hat a(\omega)-\hat a_{\rm in,1}(\omega)
\end{eqnarray}
where $\bar\theta_\Delta=\arctan\pg{2\Delta\kappa_0/\pq{\Delta^2+\kappa\pt{\kappa_1-\kappa_0}}}$ is the phase shift of the output from the first mirror with respect to the corresponding input.
Specifically the feedback works by measuring (via homodyne detection) a field quadrature of one of the two outputs of the form
\begin{eqnarray}
\hat X_{\rm out,fb}\al{\phi}(\omega)=\hat a_{\rm out,z}(\omega)\ \ee^{\ii\,\phi}+\hat a\da_{\rm out,z}(\omega)\ \ee^{-\ii\,\phi}
\end{eqnarray}
where $z=0,1$ refers to the case with feedback in reflection and in transmission, respectively.
Assuming a finite detection efficiency $\eta$, the photocurrent
is then given by
\begin{eqnarray}\label{i}
\hat i_{\rm fb}(\omega)=\sqrt{\eta}\ \hat X_{\rm out,fb}\al{\phi}(\omega)+\sqrt{1-\eta}\ \hat X_{vac}(\omega)
\end{eqnarray}
where $\hat X_{vac}(\omega)$ indicates additional vacuum noise due to non-ideal detection. Note that in the case of perfect detection efficiency
$\hat i_{\rm fb}(\omega)$ reduces to the output quadrature $\hat X_{\rm out,fb}\al{\phi}(\omega)$.

\begin{figure*}[ht!]
\begin{center}
\includegraphics[width=0.8\textwidth]{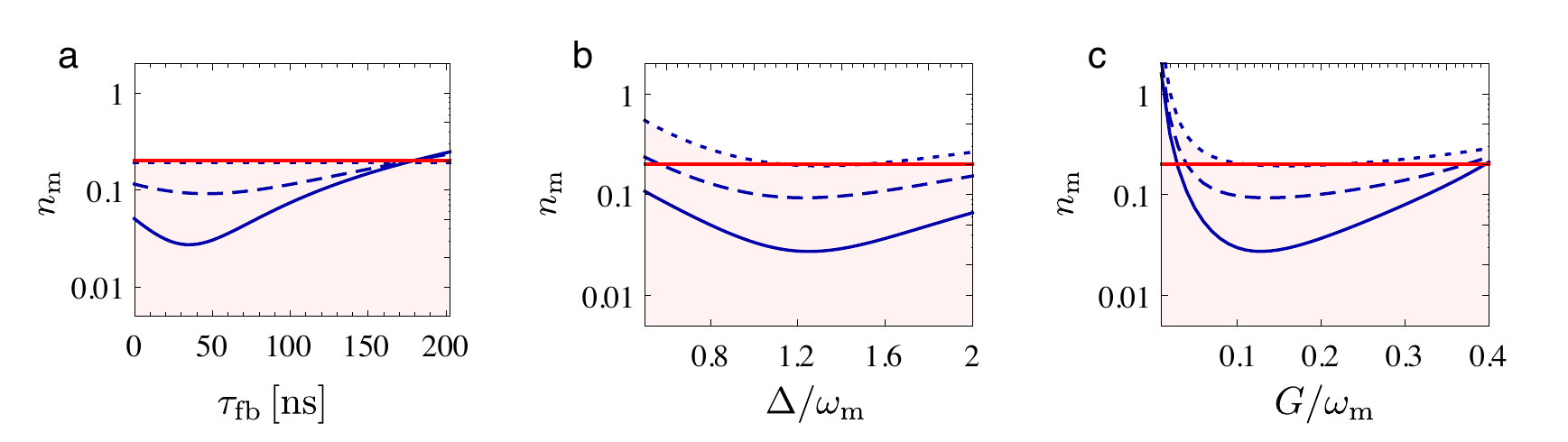}
\includegraphics[width=0.8\textwidth]{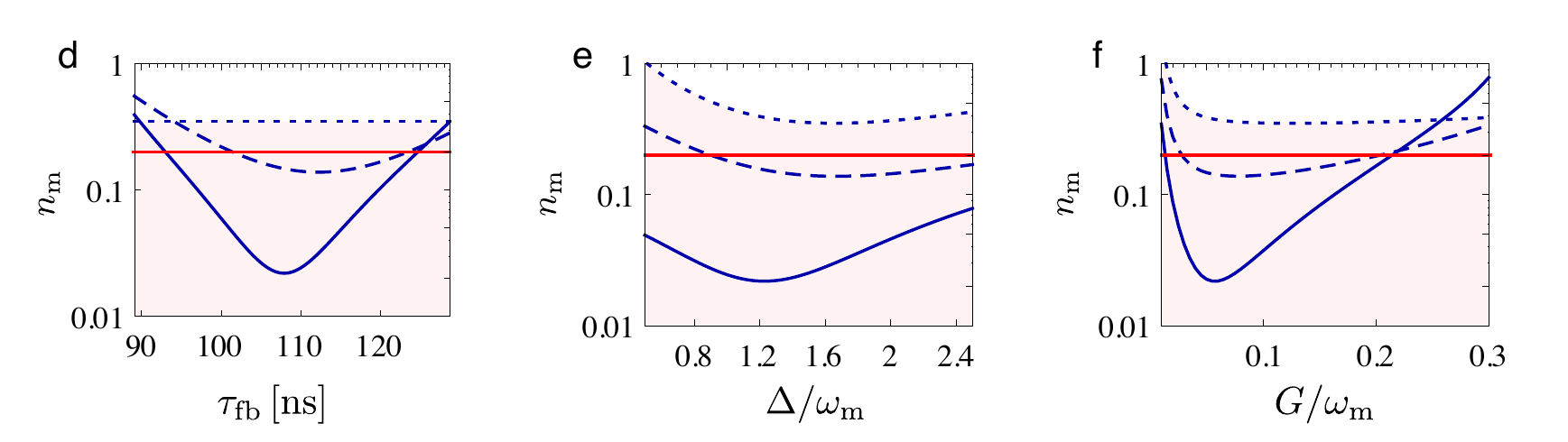}
\caption{{\bf Enhanced optical cooling in the low-noise limit.}
	Cooling with feedback-controlled light surpasses the backaction limit and is significantly stable against variation of system parameters.
	Theoretical estimate, evaluated by solving Eq.~\eqref{QLE}, for the phonon number as a function of the feedback delay time $\tau_{\rm fb}$, the cavity detuning $\Delta$ and the optomechanical coupling $G$, evaluated using the system parameters reported in
	Ref.~\cite{Clark_SM} ({\bf a}--{\bf c}) and
	Ref.~\cite{Peterson2016Laser-Cooling-o_SM} ({\bf d}--{\bf f}).
	The red lines indicate the corresponding experimental results. The light--pink area corresponds to the number of phonons beyond the backaction limit; solid--blue curves are the feedback--assisted cooling with ideal detection, dashed--blue curves correspond to efficiency $\eta = 0.42$~\cite{Purdy:2013ys_SM} in panels {\bf a}--{\bf c} and $\eta = 0.36$~\cite{Mallet:2011aa_SM} in panels {\bf d}--{\bf f}. Finally, dotted--blue curves are without feedback.
	The feedback gain $g_{\rm fb}(\omega)$ is assumed to be flat over the relevant bands of frequency and with a linear phase proportional to the feedback delay--time $\tau_{\rm fb}$, such that $g_{\rm fb}(\omega)=g_{\rm fb}\,\ee^{\ii\pt{\tau_{\rm fb}\,\omega+\pi}}$ for panels {\bf a}--{\bf c} and $g_{\rm fb}(\omega)=g_{\rm fb}\,\ee^{\ii\,\tau_{\rm fb}\,\omega}$ for panels {\bf d}--{\bf f}.
	All the lines are evaluated for the parameters which minimize the value of $n_\mathrm{m}$ as reported in the other panels and in \figurename~1 of the Letter.
	Consistently with Refs.~\cite{Clark_SM} and \cite{Peterson2016Laser-Cooling-o_SM}, the other parameters are $\omega_\mathrm{m}=2\pi\times 1.48$MHz, $\gamma_\mathrm{m}=2\pi\times 0.18$Hz,$n_T=5000$, $\kappa_0=2\pi\times1.17$MHz, $\kappa_1=0.13$MHz in panels {\bf a}--{\bf c} and $\omega_\mathrm{m}=2\pi\times 10.1$MHz, $\gamma_\mathrm{m}=2\pi\times 16$Hz,$n_T=75$, $\kappa=2\pi\times13.5$MHz, $\kappa'=50$kHz in panels {\bf d}--{\bf f}.
	Same parameters were used for obtaining the results in \figurename~1 of the Letter.
	}
\label{fig:th_SI_1}
\end{center}
\end{figure*}

Eqs.~\rp{QLE}--\rp{i} form a system of linear equations for the output and the system operators, which can be solved to determine the system steady state.

\subsection{Photocurrent}

Eqs.~\rp{QLE}--\rp{i} can be solved to determine an analytical expression for the feedback photocurrent in terms of the photocurrent with no feedback, which we indicate as $\hat i_{\rm fb}^\circ(\omega)$. Due to the linearity of the system they are proportional to each other,
\begin{eqnarray}\label{ifb}
\hat i_{\rm fb}(\omega)=\frac{1}{1-2\ \sqrt{\eta}\ \zeta_{\rm out}\al{\phi}(\omega)\ g_{\rm fb}(\omega)}\ \hat i_{\rm fb}^\circ(\omega)\ ,
\end{eqnarray}
with the proportionality factor measuring how much the light gets squashed (reduced fluctuations) or anti--squashed (enhanced fluctuations) under the effect of the feedback scheme. The function $\zeta_{\rm out}\al{\phi}(\omega)$ describes the response of the detected output quadrature to input amplitude fluctuations, namely
$\hat X_{\rm out,fb}\al{\phi}(\omega)=\zeta_{\rm out}\al{\phi}(\omega)\ \hat X_{\rm in,0}(\omega)+\zeta_{\rm out}\al{\phi+\pi/2}(\omega)\ \hat Y_{\rm in,0}(\omega)+\cdots$,
where the dots stand for additional terms proportional to other input noise operators.
The explicit form of $\zeta_{\rm out}\al{\phi}(\omega)$ is
\begin{eqnarray}\label{zetaphi}
\zeta_{\rm out}\al{\phi}(\omega)&=&\frac{\sqrt{\kappa_0\,\kappa_{\rm fb}}}{2\kappa}\pq{\chi_c(\omega)\ \ee^{\ii\pt{\phi-\theta_{\rm fb}}}+\chi_c(-\omega)^*\ \ee^{-\ii\pt{\phi-\theta_{\rm fb}}}}
\nn\\&&
- (1-z)\cos\pt{\phi-\theta_{\rm fb}}
\end{eqnarray}
where the values of the parameters $\kappa_{\rm fb}$, $\theta_{\rm fb}$ and $z$ depend on whether the feedback is based on the detection of the reflected field (output 0) or on the detection of the transmitted field (output 1): specifically, for the feedback in reflection $\kappa_{\rm fb}=\kappa_0$, $\theta_{\rm fb}=\bar\theta_\Delta$ and $z=0$, instead for the feedback in transmission $\kappa_{\rm fb}=\kappa_1$, $\theta_{\rm fb}=\theta_\Delta$ and $z=1$.
Furthermore, $\chi_c(\omega)$ is the cavity susceptibility which for an empty cavity is given by
$\chi_c(\omega)=2\kappa/\pq{\kappa+\ii\pt{\Delta-\omega}}$.
Correspondingly, the feedback photocurrent power spectrum, defined by the relation $\av{\hat i_{\rm fb}(\omega)\ \hat i_{\rm fb}(\omega')}=\delta(\omega+\omega')\ S_{i}(\omega)$, is given by the square modulus of the squashing factor
\begin{eqnarray}\label{Si}
S_{i}(\omega)=\abs{\frac{1}{1-2\ \sqrt{\eta}\ \zeta_{\rm out}\al{\phi}(\omega)\ g_{\rm fb}(\omega)}}^2\ .
\end{eqnarray}

\subsection{Cavity response to an additional seed: modified cavity susceptibility}

In order to probe the response of the cavity under the effect of the feedback, one can include an additional tone in the modulation of the AOM with an amplitude much larger than the fluctuations, but in any case much smaller than the pump field. Here we focus on the scheme based on the detection of the transmitted field, in the resolved sideband limit, which is relevant to our experiment. The corresponding photocurrent spectrum at the seed frequency [$S_{seed}(\omega_s)\propto\abs{\chi_\mathrm{c}^{\rm eff}(\omega_s)}^2$] reveals how the cavity susceptibility is modified by the feedback loop. The response function $\chi_c\al{\rm eff}$ takes the form
\begin{eqnarray}
\chi_\mathrm{c}\al{\rm eff}(\omega)=\frac{\chi_c}{1-2\,\sqrt{\eta}\ \zeta_{\rm out}\al{\phi}(\omega)\ g_{\rm fb}(\omega)}\ .
\end{eqnarray}
In the limit of small cavity linewidth $\kappa\ll\Delta$ and for a relatively short delay--time in the feedback response $\kappa\ll1/\tau_{\rm fb}$, a single pole of this function close to the cavity resonance is relevant to the system dynamics. Particularly,
in the case of a feedback system based on the detection of the transmitted field (relevant to our experiment)
it can be  approximated
(for frequency close to the cavity resonance $\omega\sim\Delta$)
as
\begin{eqnarray}\label{chieff}
\chi_c\al{\rm eff}(\omega)\simeq\frac{2\ \kappa}{\kappa_{\rm eff}+\ii\pt{\Delta_{\rm eff}-\omega}}
\end{eqnarray}
The effective cavity linewidth and detuning are
\begin{eqnarray}
\kappa_{\rm eff}&=&\kappa\left\{1-\ \Re\pq{\mathcal{T}(\Delta)\ \ee^{\ii\,\phi}}\right\}	\label{keff}\\
\Delta_{\rm eff}&=&\Delta-\kappa\Im\pq{\mathcal{T}(\Delta)\ \ee^{\ii\,\phi}} \ ,\label{Deff}
\end{eqnarray}
where $\mathcal{T}(\omega) = \sqrt{\eta\kappa_0\kappa_1}/\kappa\, g_\mathrm{fb}(\omega)\,\chi_c(\omega)\,\e^{-\im\theta_\Delta}$ is the complete open--loop transfer function, which is reported in \figurename~2(b), top. It is convenient to express these relations in terms of a normalised gain factor
\begin{eqnarray}
\GG_{\rm fb}\equiv\Re\pq{\mathcal{T}(\Delta)\ \ee^{\ii\,\phi}},
\end{eqnarray}
which is defined so that it takes the value $\GG_{\rm fb}=1$ at the feedback stability threshold, i.e. when $\kappa_{\rm eff} = 0$.
Equations~\eqref{keff}--\eqref{Deff} can be rewritten as
\begin{eqnarray}
\kappa_{\rm eff}&=&\kappa\pt{1-\GG_{\rm fb}}\\
\Delta_{\rm eff}&=&\Delta-\kappa\,\GG_{\rm fb}\,\tan\pq{\phi_{\TT}(\Delta)+\phi}\,,
\end{eqnarray}
in agreement with the expressions reported in the Letter for $\phi = 0$, and where  $\phi_\TT(\omega)$ is the phase of the open--loop transfer function $\mathcal{T}(\omega)$.

\subsection{Cooling}

\begin{figure}[ht!]
\begin{center}
\includegraphics[width=0.23\textwidth]{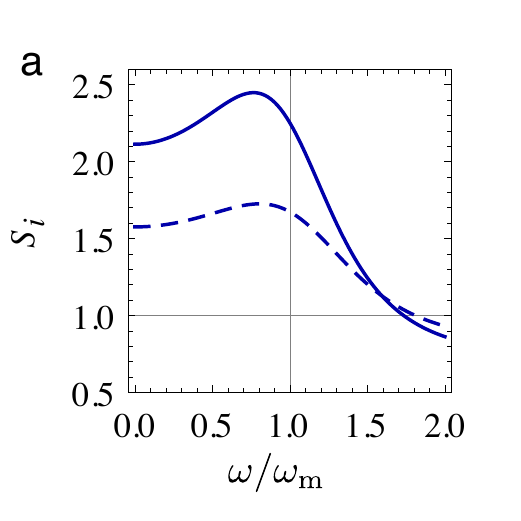}
\includegraphics[width=0.23\textwidth]{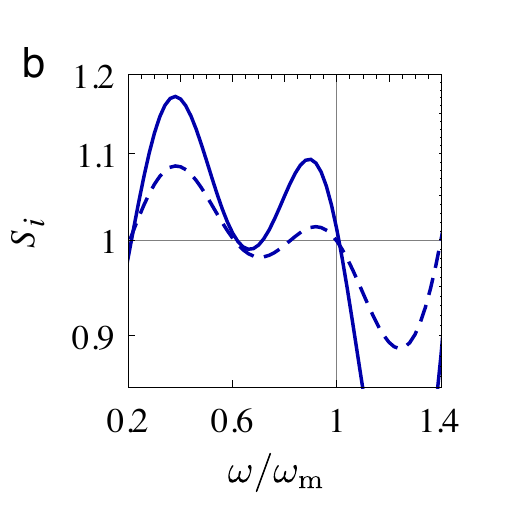}
\caption{{\bf Photocurrent power spectrum in the low--noise limit.}
	Theoretical result for the photocurrent power spectrum evaluated with Eq.~\rp{Si} in the case of an empty cavity, and with the parameters of Ref.~\cite{Clark_SM} ({\bf a}) and
	Ref.~\cite{Peterson2016Laser-Cooling-o_SM} ({\bf b}). Solid and dashed lines are for an ideal and realistic detection efficiency respectively. The feedback is tuned to achieve the minimum value of $n_\mathrm{m}$ according to the parameters used in \figurename~\ref{fig:th_SI_1} and ~\figurename~1 of the Letter.
	$S_i(\omega)=1$ corresponds to shot noise. The field fluctuations at the mechanical frequency $\omega_\mathrm{m}$ determine the strength of the optomechanical processes.
	}
\label{fig:th_SI_2}
\end{center}
\end{figure}
\begin{figure}[ht!]
\begin{center}
\includegraphics[width=0.23\textwidth]{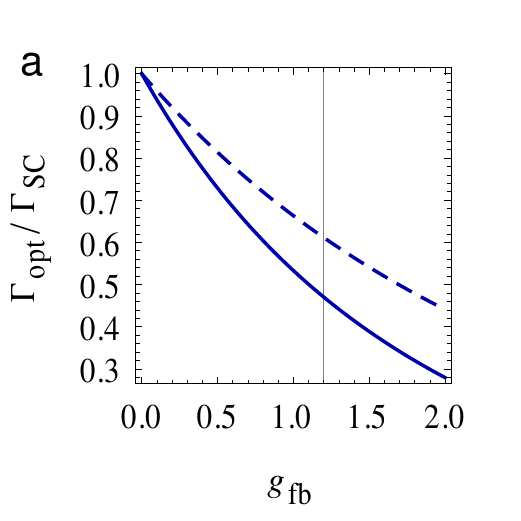}
\includegraphics[width=0.23\textwidth]{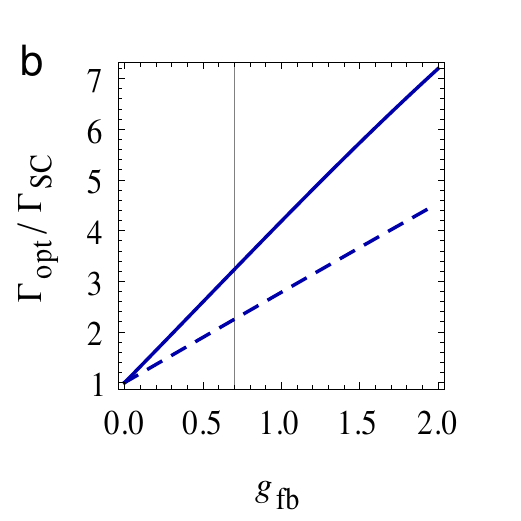}
\caption{{\bf Optomechanical cooling rate in the low--noise limit.}
	Theoretical result for the cooling rate $\Gamma_{\rm opt}=A_--A+$ (relative to the corresponding result of standard sideband cooling $\Gamma_{\rm SC}$) evaluated with Eqs.~\rp{Azero} and \rp{SX} using the parameters of Ref.~\cite{Clark_SM} ({\bf a}) and
	Ref.~\cite{Peterson2016Laser-Cooling-o_SM} ({\bf b}), and the specific parameters that minimize the value of $n_{\rm m}$ in the corresponding plots of 
	\figurename~\ref{fig:th_SI_1} and ~\figurename~1 of the Letter. Solid and dashed lines are for an ideal and realistic detection efficiency respectively. The vertical lines indicate the value of $g_{\rm fb}$ that minimize $n_{\rm m}$ as reported in~\figurename~1 of the Letter. 
	}
\label{fig:th_SI_3}
\end{center}
\end{figure}

Analytical expressions for the Stokes and anti--Stokes rate can finally be computed in the weak coupling  limit ($G\ll\omega_\mathrm{m}$). They are proportional to the power spectrum of the cavity quadrature $\hat X(\omega)=\hat a(\omega)+\hat a\da(\omega)$ evaluated for an empty cavity and at the mechanical frequency. Specifically
\begin{eqnarray}\label{Azero}
A_\pm=G^2\ S_X(\mp\omega_\mathrm{m})\ ,
\end{eqnarray}
with $\av{\hat X(\omega)\ \hat X(\omega')}=\delta(\omega+\omega')\ S_X(\omega)$.
The explicit result for the spectrum  is
\begin{eqnarray}\label{SX}
S_X(\omega)=\frac{1}{2\kappa}\lpq{
\abs{\chi(\omega)+\sqrt{\eta\frac{\kappa_{\rm fb}}{\kappa_0}} \ \Lambda(\omega)^*\ \ee^{-\ii\,\phi_{\rm fb}}}^2
}\nn\\\rpq{
+\frac{\kappa-\eta\,\kappa_{\rm fb}}{\kappa_0}\abs{\Lambda(\omega)}^2
}\ ,
\end{eqnarray}
where $\kappa_{\rm fb}=\kappa_0$ and $\phi_{\rm fb}=\phi+\theta_\Delta-\bar\theta_\Delta$ for the feedback based on the detection of the reflected field, while $\kappa_{\rm fb}=\kappa_1$ and $\phi_{\rm fb}=\phi$ for the one that measures the transmitted field.
We have also introduced the function
\begin{eqnarray}\label{Lambda}
\Lambda(\omega)&=&
\frac{2\ \zeta_c^{(0)}(\omega)\  g_{\rm fb}(\omega)}{1-2\ \sqrt{\eta}\ \zeta_{\rm out}\al{\phi}(\omega)\  g_{\rm fb}(\omega)}\ ,
\end{eqnarray}
which accounts for the effect of feedback on the mechanical processes, and
where $\zeta_c^{(0)}(\omega)$, defined by the relation $\zeta_c^{(\varphi)}(\omega)=\frac{\kappa_0}{2\kappa}
\pq{\chi_c(\omega)\ \ee^{\ii\,(\varphi-\theta_\Delta )}+\chi_c(-\omega)^*\ \ee^{-\ii\,(\varphi-\theta_\Delta)}}$, describes the response of the cavity field quadrature $\hat X(\omega)$ to input amplitude modulations, such that $\hat X(\omega)=\frac{\zeta_c^{(0)}(\omega)}{\sqrt{2\kappa_0}}\hat X_{\rm in}(\omega)+\frac{\zeta_c^{(pi/2)}(\omega)}{\sqrt{2\kappa_0}}\hat Y_{\rm in}(\omega)+\cdots$ with the dots indicating additional terms proportional to other noise operators.
When all the light lost by the cavity is detected and employed in the feedback loop (that is $\eta=1$ and a single output channel is involved), Eq.~\rp{SX} recovers Eq.~(2) of the Letter. 
In the more general case of finite detection efficiency, an additional term appears in the scattering rates, which inhibits the possibility to fully suppress Stokes scattering.
This implies that, in the low temperature limit, it is, in principle, more efficient to work in reflection with a one-sided cavity. In this case all the light lost by the cavity can, in principle, be detected and used to achieve optimal suppression of Stokes scattering.

Fig.~\ref{fig:th_SI_1} complements the theoretical findings of the Letter, obtained by solving the complete model defined by Eq.~\eqref{QLE}, in the limit of low thermal noise. It shows that, under realistic conditions, the scheme is sufficiently stable also against variations of other  parameters of the system. Fig.~\ref{fig:th_SI_2} shows that light fluctuations do not have to be suppressed altogether to achieve optimal cooling. In other words, here the feedback is not used to suppress the field noise, instead fluctuations should be properly tailored and set to specific values in order to coherently suppress unwanted heating processes. Finally, Fig.~\ref{fig:th_SI_3} shows that, in the low temperature limit, the cooling time does not change significantly with respect to standard sideband cooling.

\subsubsection*{Resolved sideband limit, high temperature limit and feedback based on the transmitted field}

Let us now analyse this result in the limit relevant to our experiment.
In the resolved sideband limit the spectrum in Eq.~\rp{SX}, evaluated for the scheme based on the detection of the transmitted field operated close to the feedback stability threshold, can be expressed in terms of the effective susceptibility defined in Eq.~\rp{chieff}.
Correspondingly, the cooling dynamics can be mapped to standard sideband cooling in an optical cavity with linewidth $\kappa_{\rm eff}$, but starting from a larger effective temperature.

Specifically we find that, in the resolved sideband limit and for frequencies close to the optical detuning, $\omega\sim\Delta$, 
\begin{eqnarray}
\Lambda(\omega)\Bigl|_{\omega\sim\Delta}\simeq \frac{\kappa_0}{\kappa}\
g_{\rm fb}(\omega)\ \chi_c\al{\rm eff}(\omega)\ \ee^{-\ii\,\theta_\Delta }
\end{eqnarray}
where we have neglected the second term in the expression for $\zeta_c\al{0}(\omega)$ which appears in the numerator of Eq.~\rp{Lambda}.
Instead for frequencies $\omega\sim-\Delta$ we find
\begin{eqnarray}
\Lambda(\omega)\Bigl|_{\omega\sim-\Delta}\simeq \frac{\kappa_0}{\kappa}\
g_{\rm fb}(\omega)\ \chi_c\al{\rm eff}(-\omega)^*\ \ee^{\ii\,\theta_\Delta }
\end{eqnarray}
where now we have neglected the first term in the expression for $\zeta_c\al{0}(\omega)$.

Correspondingly, when the cavity detuning is set at the mechanical frequency $\Delta_{\rm eff}\sim\omega_{\rm m}$ in order to optimize the anti-Stokes processes, we find that
\begin{eqnarray}
A_-&\sim&  
\frac{2\,G^2}{\kappa_{\rm eff}}+\varrho
\nn\\
A_+&\sim& 
\varrho
\label{Ap000}
\end{eqnarray}
with
\begin{eqnarray}
\varrho&=&G^2 \frac{\pt{\kappa-\kappa_{\rm eff}}^2+
\pt{\Delta-\Delta_{\rm eff}}^2
}{2\,\eta\,\kappa_1\, \kappa_{\rm eff}^2} \ 
\end{eqnarray} 
and where we have neglected additional terms proportional to $G/\omega_{\rm m}$.
In this limit the cooling rate is 
\begin{eqnarray}
\Gamma_{\rm opt}=A_--A_+\sim\frac{2\,G^2}{\kappa_{\rm eff}}\ ,
\end{eqnarray}
moreover, the parameter $\varrho$ determines the new backaction limit given by $\varrho/\Gamma_{\rm opt}$ that can be significantly larger than the standard one given instead by $ {\kappa^2}/{4\,\omega_{\rm m}^2}$. 
It is useful to express the corresponding steady state phonon number as the one achieved by standard sideband cooling in a cavity with linewidth $\kappa_{\rm eff}$. In this picture we can interpret the enhanced backaction limit as an effective enhanced temperature according to the relation   
\begin{eqnarray}\label{nm}
n_\mathrm{m} 
&\sim&n_{\rm eff}^{\rm th}\, \frac{\gamma_{\rm m}\,\kappa_{\rm eff}}{2\,G^2}
\end{eqnarray}
where the effective number of thermal excitations is 
\begin{eqnarray}\label{neff}
n_{\rm eff}^{\rm th}&=&n^{\rm th}_\mathrm{m}+\frac{\varrho}{\gamma_{\rm m}}\ .
\end{eqnarray}

We can use Eq.~\rp{nm} to express $\varrho$ as a function of the other measured quantities, namely the number of excitations of the cooled resonator $n_{\rm m}$ at \SI{0.35}{\kelvin}, the initial number of excitations $n_{\rm m}^{\rm th}$ at \SI{300}{\kelvin}, and the optomechanical coupling strength. The latter can in turn be expressed in terms of the number of excitations to which the oscillator is cooled by standard sideband cooling ($n_{\rm m}^{\rm SC}$, corresponding to \SI{2}{\kelvin}) as $G^2=n_{\rm m}^{\rm th}\ \gamma_{\rm m}\ \kappa/2\,n_{\rm m}^{\rm SC}$. Thereby we find 
\begin{eqnarray}
\varrho&=&
\gamma_{\rm m}\ n_{\rm m}^{\rm th}\
\pt{
\frac{\kappa}{\kappa_{\rm eff}}\frac{ n_{\rm m}}{n_{\rm m}^{\rm SC}}-1}\ .
\end{eqnarray}
In particular, in our experiment we can estimate that the backaction limit has been increased from an equivalent temperature of $\sim15$nK (corresponding to $\kappa^2/4\omega_{m}^2$) to $\sim 150$mK (corresponding to $\varrho/\Gamma_{\rm opt}$). Similarly, we estimate that the effective temperature, defined in Eq.~\rp{neff}, has been enhanced by roughly $\sim 225$K (corresponding to $\varrho/\gamma_{\rm m}$). 

We can also rewrite Eq.~\rp{nm} in order to express how much the phonon number is reduced with respect to the standard sideband cooling result $n_{\rm m}^{\rm SC}$. We find
\begin{eqnarray}\label{nmeff}
n_\mathrm{m} &\sim& n_{\rm m}^{\rm SC}\ \frac{\kappa_{\rm eff}}{\kappa}+
\frac{\pt{\kappa-\kappa_{\rm eff}}^2+\pt{\Delta-\Delta_{\rm eff}}^2}{4\,\eta\,\kappa_1\,\kappa_{\rm eff}}
\ .
\end{eqnarray}
When $\Delta=\Delta_{\rm eff}$, this value can always be made smaller than $n_{\rm m}^{\rm SC}$. Specifically, the minimum is found for $\kappa_{\rm eff}=\kappa\,\sqrt{\kappa/\pt{4\,n_{\rm m}^{\rm SC}\,\eta\,\kappa_1+\kappa}}$, and the corresponding minimum value is 
\begin{eqnarray}\label{nmmin}
n_{\rm m}\sim\frac{2 \, n_{\rm m}^{\rm SC}}{1 + \sqrt{1 + 4\,\eta\,\kappa_1 \, n_{\rm m}^{\rm SC}/\kappa}} \, ,
\end{eqnarray}
which is always strictly smaller than $n_{\rm m}^{\rm SC}$. 
In particular, in the limit of high temperature, for perfect detection efficiency, and for a symmetric cavity, the minimum number of excitations is
\begin{eqnarray}
n_{\rm m}\sim\sqrt{2 \,n_{\rm m}^{\rm SC}}\ .
\end{eqnarray}

\section{\label{sec:system_char} Experiment}

\begin{figure}[ht!]
\begin{center}
\includegraphics[width=.45\textwidth]{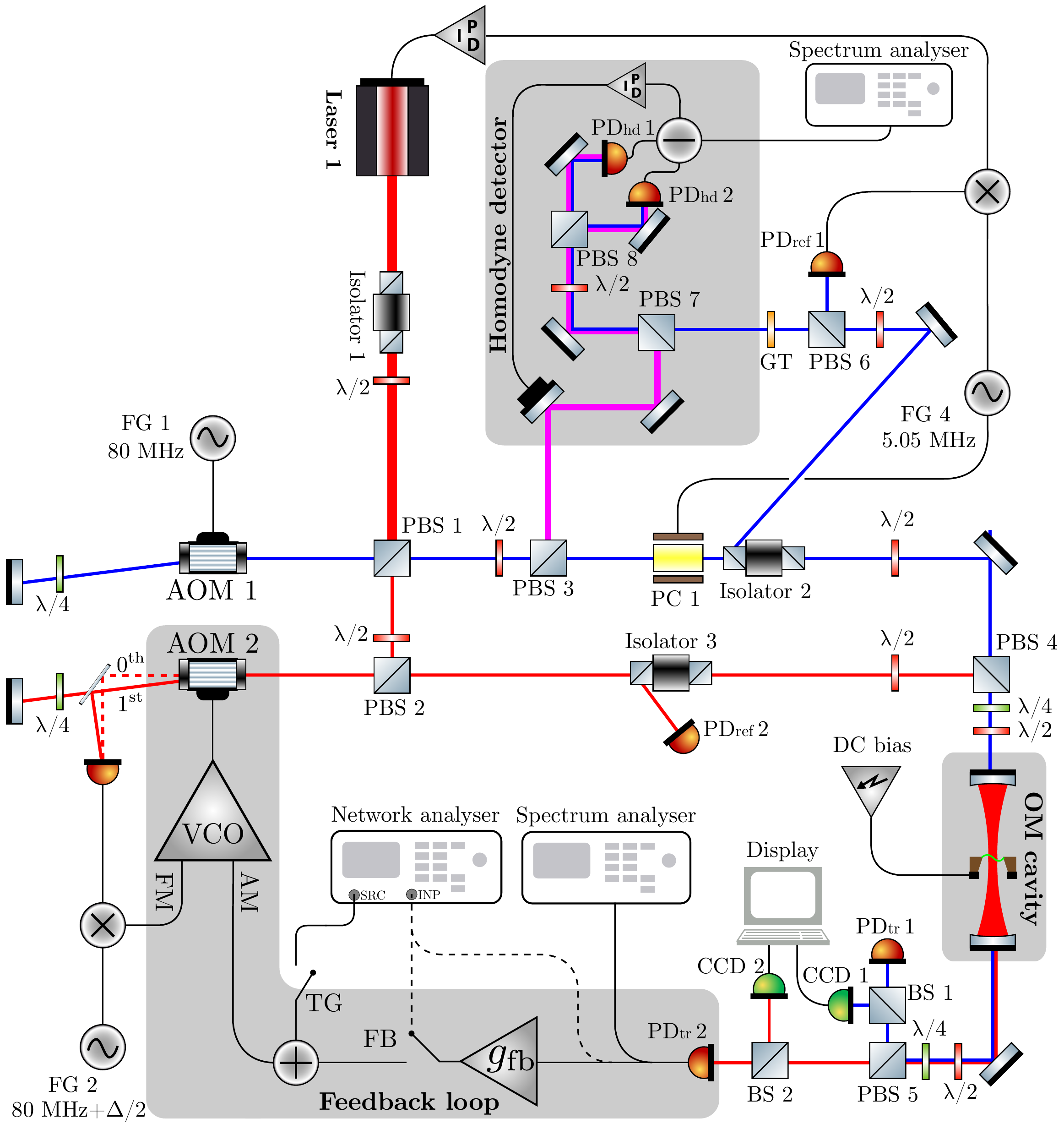}
\caption{{\bf A schematic of the experimental setup}.}
\label{fig:expsetup}
\end{center}
\end{figure}

A detailed experimental scheme is reported in~\figurename~\ref{fig:expsetup}. The frequencies of a probe (blue line) and a cooling (red line) beam, derived from a \SI{1064}{\nano\meter}~master laser, are shifted by two acousto--optic modulators (AOMs) in a double passage configuration.
The AOM on the probe beam, held on resonance with the cavity by means of a Pound--Drever--Hall locking scheme~\cite{Black2001An-introduction_SM}, is always driven at \SI{80}{\mega\hertz}.
The membrane displacement is revealed by detecting the phase of the reflected probe beam employing a balanced homodyne detector~\cite{Yuen1983Noise-in-homody_SM}.
The orthogonally polarised probe and local oscillator (purple line) beams are combined and then split by two sets constituted of a waveplate and a polarising beam--splitter.
A detuning $\Delta$ is imparted on the cooling beam by means of a commercial VCO, which is controlled by a phase--locked--loop (PLL), derived by demodulating the beating tone between the $0^\mathrm{th}$ and $1^\mathrm{st}$ order diffracted beams at the frequency of $\SI{80}{\mega\hertz}+\Delta/2$ (the divisor 2 being due to double passage through the AOM).
A feedback loop on the cooling field is implemented by driving the VCO amplitude modulation input (AM) with the amplified and filtered ($g_\mathrm{fb}$) electronic signal acquired by detecting the light intensity either transmitted or reflected by the cavity. Two sets of waveplates put in front of each cavity mirror allow one to split the transmitted cooling and probe beams with a high extinction ratio
by exploiting orthogonal polarisations.
The open--loop transfer functions of the feedback, i.e. the curves reported in the top of panel b) in~\figurename~2 of the Letter, are measured by closing the switch TG on the network analyser tracking generator, which provides a frequency swept tone to the VCO input, adding in turn a seed on the cooling beam; the electronic signal is analysed with the switch FB open.
The closed--loop transfer function, that is the effective cavity response in~\figurename~3 of the Letter, is instead measured by closing the switch FB and analysing directly the photocurrent.
The amplitude noise fluctuations, i.e. the curves reported in the bottom of~\figurename~2(b) of the Letter, are revealed by analysing directly the photocurrent with the switch TG open and FB closed. To calibrate the feedback gain $\mathcal{G}_\mathrm{fb}$, we increase the filter gain up to the loop instability, i.e. $\mathcal{G}_\mathrm{fb}\equiv1$ and then added commercial calibrated attenuators.
Cavity and membrane are both embedded in a vacuum chamber operating at $\SI{7e-7}{\milli\bar}$.

\subsection{\label{sec:cavity_response}Cavity decay rate}

\begin{figure}[ht!]
\begin{center}
   {\includegraphics[width=.45\textwidth]{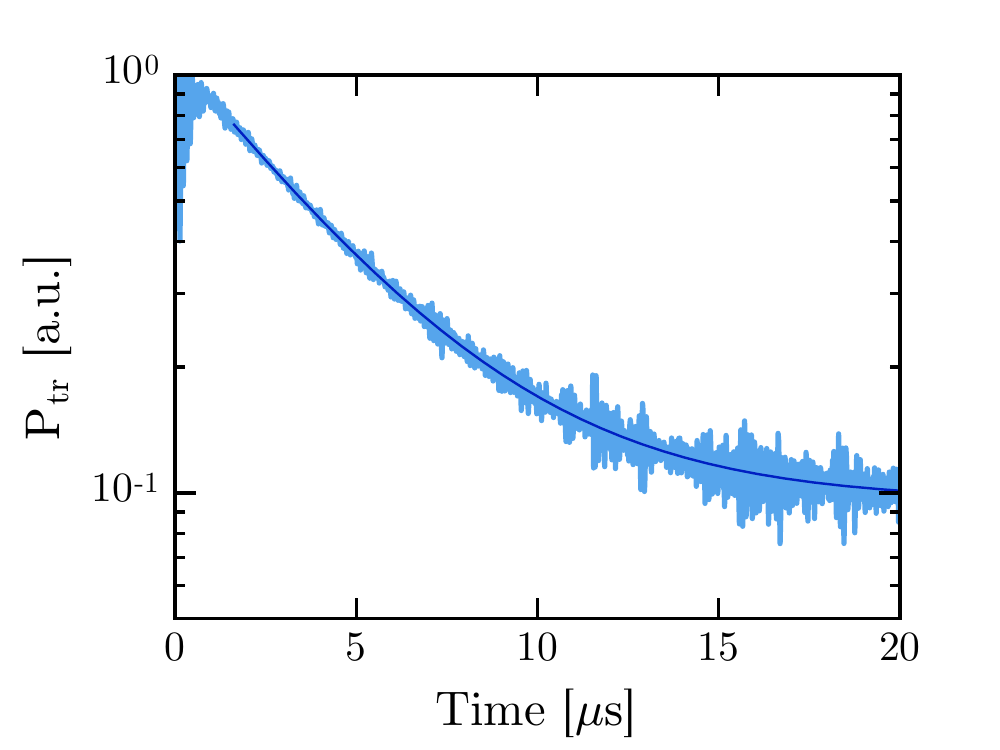}}
 \caption{{\bf Cavity decay rate measurement}. The exponential decay of the light transmitted by the cavity is fitted with a decay rate $2\kappa = 2\pi\times$\SI{40.3}{\kilo\hertz}.}
\label{fig:cavity_char}
\end{center}
\end{figure}

The cavity decay rate was characterised by employing a ring--down technique, i.e. measuring the exponential decay of the light intensity leaking out from the cavity, as shown in \figurename~\ref{fig:cavity_char}. Here the membrane is placed in a cavity field node to rule out the optomechanical coupling. The fitted
decay rate is $2\kappa = 2\pi\times$\SI{40.3}{\kilo\hertz}. An independent measurement obtained by scanning the probe beam frequency and fitting the data with a lorentzian lineshape results in $2\kappa=2\pi\times$ \SI{40.2}{\kilo\hertz}, in accordance with the time measurement.
In general, the cavity decay rate depends on the membrane position along the cavity~\cite{Serra2016aa_SM}. For the optomechanical experiment reported in the Letter the membrane position determines a cavity decay rate $2\kappa = 2\pi\times$\SI{43}{\kilo\hertz}.

\subsection{\label{sec:cal_g0}Single--photon coupling calibration}

The calibration of the single--photon optomechanical coupling $g_0$ is obtained by following the technique employed in Ref.~\cite{Gorodetksy2010Determination-o_SM}.
An external calibration monotone with frequency near the mechanical one ($\omega_{\mathrm{cal}} = 2\pi\times$\SI{340.8}{\kilo\hertz}), is added to the phase modulation of the probe beam used for locking ($\omega_{\mathrm{PDH}} = 2\pi\times$\SI{5.05}{\mega\hertz}) by means of a Pockels cell.
\begin{figure}[ht!]
\begin{center}
\includegraphics[width=.45\textwidth]{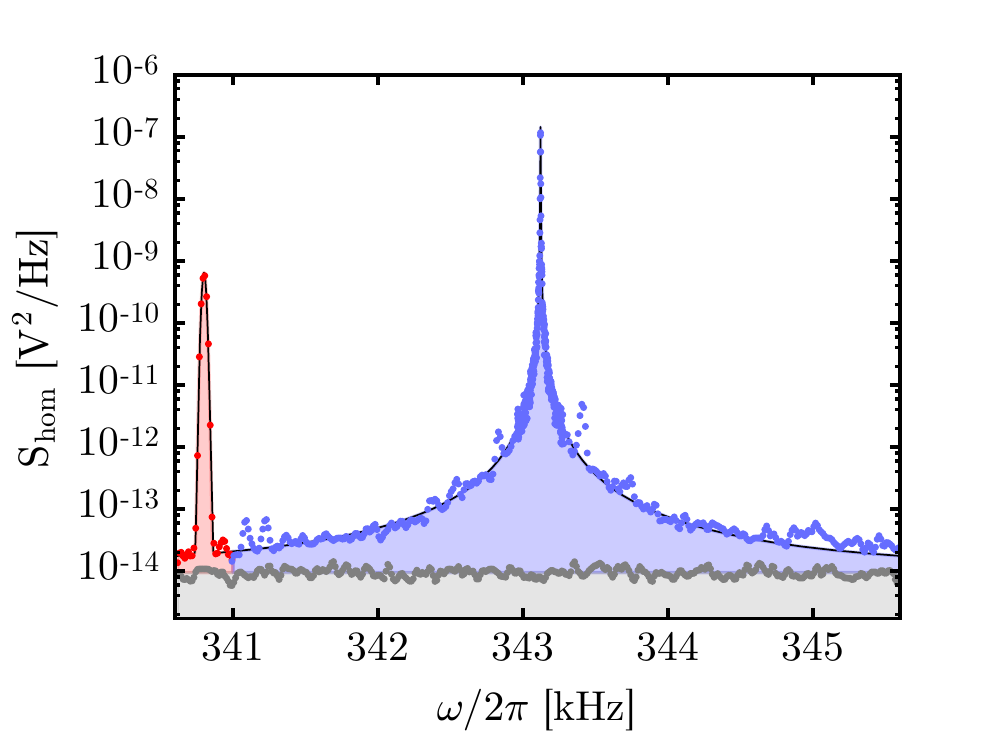}
\caption{{\bf Optomechanical coupling rate calibration}.
Typical spectrum of the homodyne photocurrent $S_\mathrm{hom}$. The blue symbols are due to the thermal fluctuations of the fundamental mechanical mode, fitted by a lorentzian curve (light--blue area). The leftmost gaussian feature, fitted by the light--red area, is the calibration tone at $\omega_{\mathrm{cal}} = 2\pi\times$\SI{340.8}{\kilo\hertz}. The ratio of these areas is used to calibrate the optomechanical coupling rate $g_0$.}
\label{fig:calibration_go}
\end{center}
\end{figure}
A typical homodyne result of the thermal noise of the fundamental mode of the circular membrane  is shown in~\figurename~\ref{fig:calibration_go}. We identify two main contributions in the spectrum: the blue symbols are due to displacement fluctuations of the fundamental mechanical mode, while the red ones account for the calibration tone. The single--photon optomechanical coupling $g_0$ is estimated by comparing the area of the two contributions, by knowing the temperature, and by properly calibrating the Pockels cell modulation depth, $\beta$. The modulation depth is independently calibrated by means of a heterodyne technique, showing a linear behaviour $\beta\,[\mathrm{mrad}] \approx 10^{-2} V_{\mathrm{mod}}\,[\mathrm{mVpp}]$ in the range 10--\SI{e3}{\milli\volt pp}. The amplitude modulation of the calibration tone used for estimating $g_0$, as in~\figurename~\ref{fig:calibration_go}, was \SI{500}{\milli\volt pp}. The single--photon optomechanical coupling amounts to $g_0 = 2\pi\times$\SI{0.84}{\hertz}. The optomechanical coupling $G$ from the Letter is related to $g_0$ by the relation $G = g_0\sqrt{2}n_\mathrm{c}$, where $n_\mathrm{c} = 2\kappa_0\mathcal{P}/\hbar \omega_\mathrm{L}(\kappa^2 + \Delta^2)$ is the mean photon number in the cavity, with $\mathcal{P}$ the input power of the cooling beam.

\subsection{Circular membrane normal modes}\label{sec:vibrations_mem}

For a circular membrane with a radius $R$, the transverse deformation functions of normal modes, neglecting bending effects for a high--stress taut membrane, are a product of the radial and angular contributions~\cite{ElmorePhysicsWave_SM}
\begin{eqnarray}
u^\mathrm{m}_{\mathrm{nj}}(r,\theta)^\pm=J_\mathrm{n}\left(\alpha_\mathrm{nj}\frac{r}{R}\right)\cos\left(\mathrm{n}\theta\pm\phi_\pm^{(n)}\right),
\end{eqnarray}
where $\mathrm{n}=0,1,2\dots$, $\mathrm{j}=1,2\dots$, $J_\mathrm{n}$ is the n--th Bessel function of the first kind, and $\alpha_{\mathrm{nj}}$ the \textit{j}--th zero of $J_\mathrm{n}$. The normal mode $\mathrm{(n,j)}$ oscillates in time with eigenfrequency
\begin{eqnarray}
\omega_\mathrm{m}^\mathrm{nj} = \frac{c_s}{R}\alpha_\mathrm{nj},\qquad c_s=\sqrt{\frac{T}{\rho}},
\end{eqnarray}
where $c_s$ is the sound velocity in the material, $T$ is the surface tension and $\rho$ is the mass density.
Apart from the case $\mathrm{n}=0$, all the eigenfrequencies are doubly degenerate: the two phases $\phi_\pm^{(n)}$ are nonzero and arbitrary, satisfying the condition $\phi_+^{(n)}-\phi_-^{(n)}=\pi/2$, which guarantees the orthogonality of the modes.

The effective mass is another important property which differs for circular membranes as compared to the square ones~\cite{Serra2016aa_SM}. Generally speaking, the mechanical oscillator displacement should be addressed with a three--dimensional vector. However, it is usual in optomechanics to map the vibrational motion in a scalar quantity, $x(t)$. In doing this, the normalisation of the normal mode functions $u^\mathrm{m}_\mathrm{kj}(x,y)$ can be arbitrarily chosen, leading to an ambiguity in the definition of the mass. Nevertheless, independently of the normalisation adopted, the potential energy stored in the oscillator must be $U = \pt{m_{\mathrm{eff}}/2}\omega_\mathrm{m}^2\,x(t)^2$. This relation establishes a way to define an effective mass. For a circular membrane we have
\begin{align}
m_{\mathrm{eff}}^\mathrm{nj} = \rho L_\mathrm{d} \pi R^2\int_0^1 xJ_\mathrm{n}(\alpha_\mathrm{nj}x)^2dx = m \int_0^1 xJ_\mathrm{n}(\alpha_\mathrm{nj}x)^2dx,
\end{align}
where $m$ is the physical mass of the resonator. For $\mathrm{n}=0$, we have $m^\mathrm{0j}_{\mathrm{eff}} = m J_1^2(\alpha_\mathrm{0j})$.

The characterisation of normal modes of the circular membrane used in the experiment, with radius \SI{0.615}{\milli\meter} and thickness \SI{97}{\nano\meter}, is made by putting it in the middle of the optical cavity and monitoring, with a balanced homodyne detector, the phase of the reflected probe beam.
Due to its low optical power and to the resonant condition, dynamical backaction from this beam can safely be neglected.
\begin{figure}[ht!]
\begin{center}
   {\includegraphics[width=.45\textwidth]{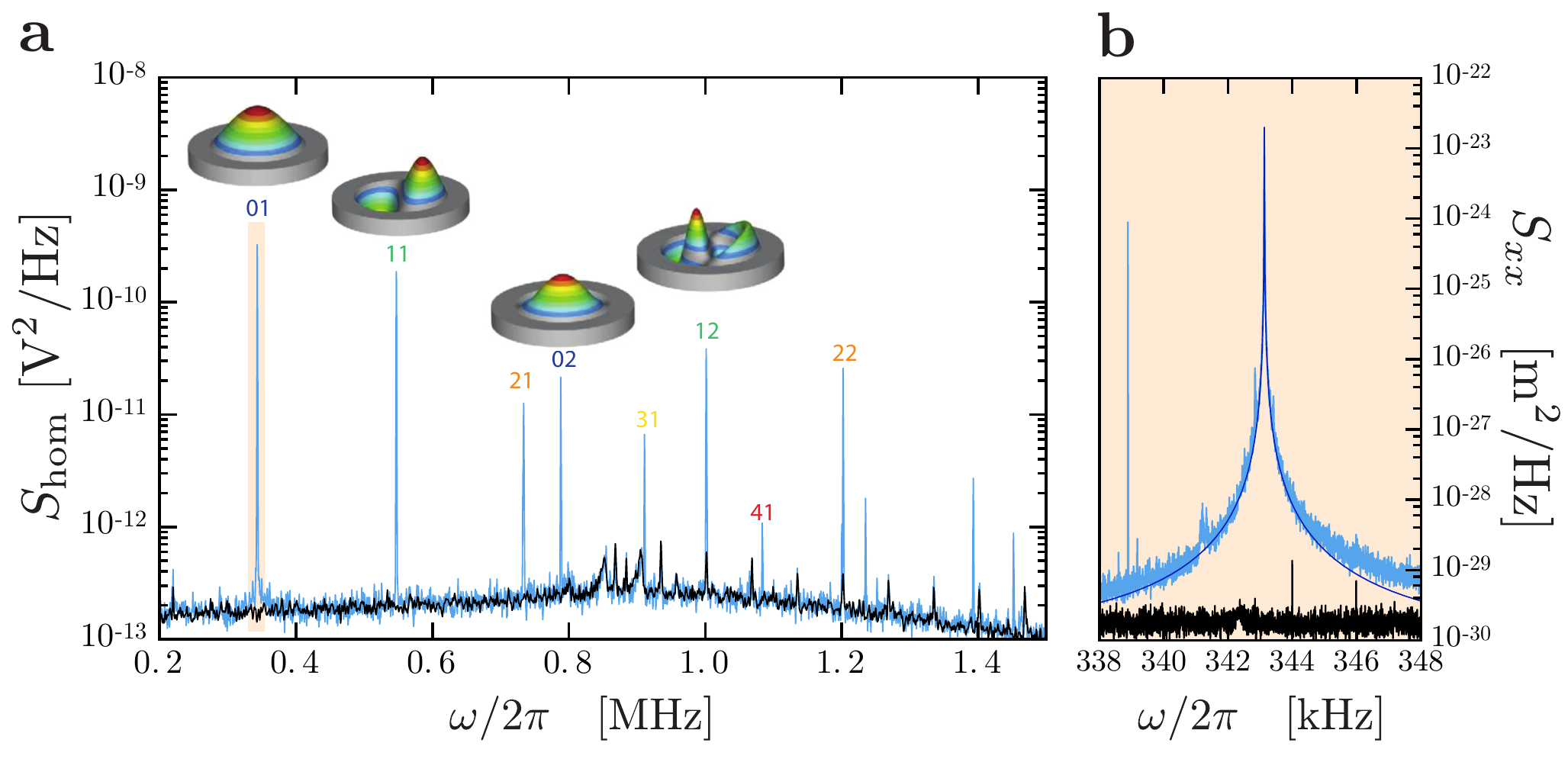}}
 \caption{{\bf Characterisation of the mechanical oscillator}.
	{\bf a}, Broadband homodyne signal spectrum expressed as apparent output voltage noise. The solid blue line corresponds to the measurement data, and the black one to shot--noise. The colored orange area highlights the fundamental mode of the membrane. Noise peaks arising from the membrane thermal motion are labeled by mode numbers ${\rm (n,j)}$. In particular, four transverse deformation functions are pictorially represented above each corresponding mode.
	{\bf b}, Thermal noise displacement spectrum $S_{xx}$ of the fundamental mode (0,1). The black line represents the shot--noise; the blue line corresponds to measurement data; the dark--blue line corresponds to the estimated contribution from thermal noise at \SI{300}{\kelvin}. Fitting the data with a theoretical power spectral density of the position of a mechanical oscillator undergoing brownian motion, a resonance frequency $\omega_\mathrm{m} \simeq 2\pi \times\SI{343.13}{\kilo\hertz}$, and decay rate $\gamma_\mathrm{m}\simeq 2\pi\times\SI{1.18}{\hertz}$ are estimated.}
\label{fig:membrane_char}
\end{center}
\end{figure}
In \figurename~\ref{fig:membrane_char}(a) a broadband spectrum is reported. The black trace is the detection shot--noise. In the light--blue trace, several peaks are present and they can be recognised as the normal modes [labeled by (n,j)] of the circular membrane put in the middle of the cavity. To characterise the fundamental mode (0,1), which is used throughout the work, we zoomed in, as shown in {\bf b}, and fitted the resulting lineshape, estimating a resonance frequency $\omega_\mathrm{m} \simeq 2\pi \times\SI{343.13}{\kilo\hertz}$ and decay rate $\gamma_\mathrm{m}\simeq 2\pi\times\SI{1.18}{\hertz}$.
The leftmost narrow feature in \figurename~\ref{fig:membrane_char}(b) is the calibration tone used for estimating $g_0$.

\subsection{\label{sec:filter_loop} Decomposition of the open--loop response}

By exploiting orthogonal polarisations, the transmitted cooling and probe beams are split with high extinction ratio. The total open--loop transfer function $\mathcal{T}(\omega)$ is measured by collecting the light of the cooling beam with a single InGaAs photodiode and converting the photocurrent into a voltage signal by means of a transimpedance amplifier. The network analysis is performed on the signal after a high--pass electronic filter, a PD controller, with a corner frequency of $\SI{150}{\kilo\hertz}$.
\begin{figure}[h!]
\includegraphics[width=0.45\textwidth]{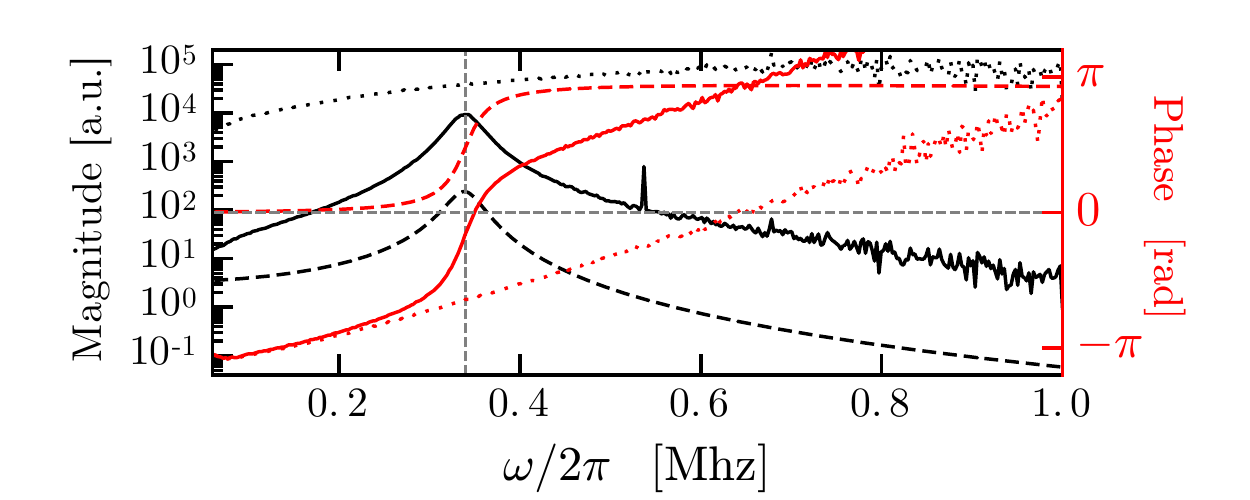}
\caption{\textbf{Transmission feedback open--loop response.}
	The feedback loop with transmitted light contains both optical and electronic filters. The complete measured open--loop response $\mathcal{T}(\omega)$ contains both the optical and the electronic transfer functions. In order to properly assess the theoretical analysis, we decompose it in the two parts. Black and red lines correspond, respectively, to magnitude and phase of the complex functions. Solid lines are the data measured; dashed lines are estimates of the cavity transfer function; dotted lines are the electronic filter, obtained by subtracting the cavity function from the data. A vertical grey--dashed line represents the detuning of the cooling beam.}
\label{fig:transmission_squashing_SI}
\end{figure}
In \figurename~\ref{fig:transmission_squashing_SI}, black and red solid lines represent, respectively, the magnitude and phase of the complete open--loop transfer functions.
In order to properly characterise the feedback, the measured open--loop response is decomposed in the cavity and the electronic transfer function.
The cavity transfer function (dashed lines) inferred from the measured decay rate (see \ref{sec:cavity_response}) and the detuning, is subtracted from the complete open--loop response, leaving the electronic filter transfer function (dotted lines)
used for obtaining the theoretical effective temperature.
$\tau_\mathrm{fb}$ is estimated from the slope of the electronic feedback phase to be $\SI{750}{\nano\second}$.

\subsection{\label{sec:cooling} Cooling}

\begin{figure}[h!]
\includegraphics[width=0.3\textwidth]{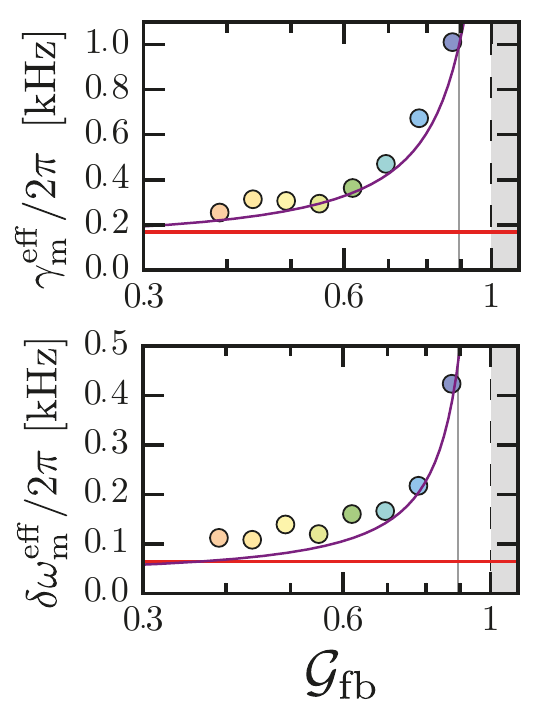}
\caption{\textbf{Feedback effects on the mechanical susceptibility.} The effective mechanical decay rate $\gamma_\mathrm{m}^\mathrm{eff}$ ({\bf a}) and resonance shift $\delta\omega_\mathrm{m}^\mathrm{eff}\equiv\omega_\mathrm{m}^\mathrm{eff}-\omega_\mathrm{m}$ ({\bf b}) increase with the normalized feedback gain $\mathcal{G}_\mathrm{fb}$. The two data--points of each colour (corresponding to $\gamma_\mathrm{m}^\mathrm{eff}$ and $\delta\omega_\mathrm{m}^\mathrm{eff}$ respectively) are given by the best Lorentzian fit of the spectrum of matching colour in \figurename~4(a) of the Letter. The solid purple and red lines are theoretical predictions with and without feedback respectively. The grey area represents the instability region, and the vertical grey lines indicate the optimal gain value for cooling.
	}
\label{fig:mech_shift_broadening_SI}
\end{figure}

As visible from the spectra in \figurename~4(a) of the Letter, the mechanical susceptibility is broadened and shifted as the feedback loop is closed in transmission and the gain is increased. In \figurename~\ref{fig:mech_shift_broadening_SI} we quantify these effects by plotting explicitly the effective mechanical decay rate $\gamma_\mathrm{m}^\mathrm{eff}$ and resonance shift $\delta\omega_\mathrm{m}^\mathrm{eff}\equiv\omega_\mathrm{m}^\mathrm{eff}-\omega_\mathrm{m}$, inferred from the best fits of the said spectra.

The effective mechanical energy [cf. \figurename~4(b)~and~(c) of the Letter], i.e. the effective temperature, is also inferred from the noise displacement spectra. Firstly, the noise displacement fluctuations of the fundamental mode of the mechanical oscillator, at room temperature, are recorded blocking the cooling beam and turning off the feedback.
Since the probe beam is weak and resonant, we assume that there are no radiation pressure effects.
Knowing the single--photon coupling $g_0$, and assuming an equilibrium temperature of \SI{300}{\kelvin} and an effective mass of \SI{48.2}{\nano\gram}, the spectra are calibrated in displacement units.
Once the radiation--pressure interaction and the feedback are restored, assuming the validity of the equipartition theorem~\cite{Genes2008Ground-state-co_SM}, the effective temperature is inferred from the displacement variance, evaluated by a numerical integration of the calibrated measured spectra, after shot--noise subtraction (which is always 10 dB below each spectrum).
The temperature is reduced by a factor given by the ratio of the estimated variances and the one obtained without radiation--pressure interaction and feedback.


\begin{thebibliography}{40}

\bibitem{WisemanBook}
H. M. Wiseman and G. J. Milburn,
\newblock{{\it Quantum Measurement and Control}}
\newblock{(Cambridge University Press, Cambridge, 2010)}

\bibitem{Sayrin2011realtime}
C. Sayrin {\it et al.},
\newblock{{Nature} {\bf 477}, 73 (2011).}

\bibitem{Riste2013}
D. Rist\'e {\it et al.},
\newblock{{Nature} {\bf 502}, 350 (2013).}

\bibitem{Wilson2015Measurement-bas}
D. J. Wilson {\it et al.},
\newblock{{Nature} {\bf 524}, 325 (2015).}

\bibitem{Taubman1995Intensity-feedb}
M. S. Taubman {\it et al.},
\newblock{{J. Opt. Soc. Am. B} {\bf 12}, 1792 (1995).}

\bibitem{Wiseman1999Squashed-states}
H. M. Wiseman,
\newblock{{J. Opt. B: Quantum Semiclass. Opt.} {\bf 1}, 459 (1999).}

\bibitem{Shapiro1987Theory-of-light}
J. H. Shapiro {\it et al.},
\newblock{{J. Opt. Soc. Am. B} {\bf 4}, 1604 (1987).}

\bibitem{Wiseman1998In-Loop-Squeezi}
H. M. Wiseman,
\newblock{{Phys. Rev. Lett.} {\bf 81}, 3840 (1998).}

\bibitem{Buchler1999Suppression-of-}
B. C. Buchler {\it et al.},
\newblock{{Opt. Lett.} {\bf 24}, 259 (1999).}

\bibitem{Sheard2005Experimental-de}
B. S. Sheard {\it et al.},
\newblock{{IEEE J. Quan. Elect.} {\bf 41}, 434 (2005).}

\bibitem{Cohadon1999}
P. F. Cohadon {\it et al.},
\newblock{{Phys. Rev. Lett.} {\bf 83}, 3174 (1999).}

\bibitem{Poggio2007Feedback-Coolin}
M. Poggio {\it et al.},
\newblock{{Phys. Rev. Lett.} {\bf 99}, 017201 (2007).}

\bibitem{Vinante2008Feedback-Coolin}
A. Vinante {\it et al.},
\newblock{{Phys. Rev. Lett.} {\bf 101}, 033601 (2008).}

\bibitem{McKenzie2002Experimental-De}
K. McKenzie {\it et al.},
\newblock{{Phys. Rev. Lett.} {\bf 88}, 231102 (2002).}

\bibitem{LIGO2013}
J. Aasi {\it et al.},
\newblock{{Nat. Photon.} {\bf 7}, 613 (2013).}

\bibitem{Peano2015Intracavity-Squ}
V. Peano {\it et al.},
\newblock{{Phys. Rev. Lett.} {\bf 115}, 243603 (2015).}

\bibitem{Clark2016Observation-of-}
J. B. Clark {\it et al.},
\newblock{{Nat. Phys.} {\bf 12}, 683 (2016).}

\bibitem{Schafermeier2016aa}
C. Sch{\"a}fermeier {\it et al.},
\newblock{{Nat. Commun.} {\bf 7}, 13628 (2016).}

\bibitem{Clark}
J. B. Clark {\it et al.},
\newblock{{Nature} {\bf 541}, 191 (2017).}

\bibitem{Asjad2016Suppression-of-}
M. Asjad {\it et al.},
\newblock{{Phys. Rev. A} {\bf94}, 051801 (2016).}

\bibitem{AspelmeyerRMP}
M. Aspelmeyer {\it et al.},
\newblock{{Rev. Mod. Phys.} {\bf 86}, 1391 (2014).}

\bibitem{Teufel2011}
J. D. Teufel {\it et al.},
\newblock{{Nature} {\bf 475}, 359 (2011).}

\bibitem{Chan2011}
J. Chan {\it et al.},
\newblock{{Nature} {\bf 478}, 89 (2011).}

\bibitem{Peterson2016Laser-Cooling-o}
R. W. Peterson {\it et al.},
\newblock{{Phys. Rev. Lett.} {\bf116}, 063601 (2016).}

\bibitem{SM}
See the Supplemental Material at URL for a comprehensive description of the mathematical model at the basis of our investigation, and for details regarding the experimental setup and the analysis of the experimental data, which includes 
Refs.~\cite{Genes2008, Clark, Peterson2016Laser-Cooling-o, Purdy:2013ys, Mallet:2011aa, Black2001, Yuen1983, Serra2016Microfabricatio, Gorodetksy2010, ElmorePhysicsWave}.

\bibitem{Karuza:2012fk}
M. Karuza {\it et al.},
\newblock{{New J. Phys.} {\bf14}, 095015 (2013).}

\bibitem{Karuza2013aa}
M. Karuza {\it et al.},
\newblock{{Phys. Rev. A} {\bf88}, 013804 (2013).}

\bibitem{Thompson:2008uq}
J. D. Thompson {\it et al.},
\newblock{{Nature} {\bf 452}, 72 {2008}.}

\bibitem{Serra2016Microfabricatio}
E. Serra {\it et al.},
\newblock{{AIP Adv.}  {\bf 6}, 065004 (2016).}

\bibitem{Purdy:2013ys}
T. P. Purdy {\it et al.},
ì\newblock{{Phys. Rev. X} {\bf 3}, 031012 (2013).}

\bibitem{Mallet:2011aa}
F. Mallet {\it et al.},
\newblock{{Phys. Rev. Lett.} {\bf 106}, 220502 (2011).}

\bibitem{Genes2008}
C. Genes {\it et al.},
\newblock{{Phys. Rev. A} {\bf 77}, 033804 (2008).}

\bibitem{Black2001}
E. D. Black,
\newblock{{Am. J. Phys.} {\bf 69}, 79--87 (2001).}

\bibitem{Yuen1983}
H. P. Yuen, and V. W. S. Chan,
\newblock{{Opt. Lett.} {\bf 8}, 177--179 (1983).}

\bibitem{Gorodetksy2010}
M. L. Gorodetksy {\it et al.},
\newblock{{Opt. Express} {\bf18}, 23236--23246 (2010).}

\bibitem{ElmorePhysicsWave}
W. C. Elmore and M. A. Heald,
\newblock{{\it Physics of waves.}}
\newblock{(Dover, 1985).}

\end{thebibliography}

\begin{thebibliography}{99}



\bibitem{Genes2008Ground-state-co_SM}
C. Genes, D. Vitali, P. Tombesi, S. Gigan, and M. Aspelmeyer.
\newblock{{\it Phys. Rev. A} {\bf 77}, 033804 (2008).}

\bibitem{Clark_SM}
J. B. Clark, F. Lecocq, R. W. Simmonds, J. Aumentado, and J. D. Teufel.
\newblock{{\it Nature} {\bf 541}, 191--195 (2017).}

\bibitem{Peterson2016Laser-Cooling-o_SM}
R. W. Peterson {\it et al.}
\newblock{{\it Phys. Rev. Lett.} {\bf116}, 063601 (2016).}

\bibitem{Purdy:2013ys_SM}
T. P. Purdy, P. L. Yu, R. W. Peterson, N. S. Kampel, and C. A. Regal,
\newblock{{\it Phys. Rev. X} {\bf 3}, 031012 (2013).}

\bibitem{Mallet:2011aa_SM}
F. Mallet {\it et al.}
\newblock{{\it Phys. Rev. Lett.} {\bf106}, 220502 (2011).}

\bibitem{Black2001An-introduction_SM}
E. D. Black,
\newblock{{\it Am. J. Phys.} {\bf 69}, 79--87 (2001).}

\bibitem{Yuen1983Noise-in-homody_SM}
H. P. Yuen, and V. W. S. Chan,
\newblock{{\it Opt. Lett.} {\bf 8}, 177--179 (1983).}

\bibitem{Serra2016aa_SM}
E. Serra {\it et al.}
\newblock{{\it AIP Adv.} {\bf 6}, 065004 (2016).}

\bibitem{Gorodetksy2010Determination-o_SM}
M. L. Gorodetksy, A. Schliesser, G. Anetsberger, S. Deleglise, and T. J. Kippenberg,
\newblock{{\it Opt. Express} {\bf18}, 23236--23246 (2010).}

\bibitem{ElmorePhysicsWave_SM}
W. C. Elmore and M. A. Heald,
\newblock{{\it Physics of waves.}}
\newblock{(Dover, 1985).}

\end{thebibliography}
\end{document}